\documentclass[a4paper,11pt]{article}
\pdfoutput=1
\usepackage{jcappub}

\usepackage{tikz,xcolor,hyperref}
\usepackage{amsmath, amssymb, amsthm, graphicx, epsfig, fancyhdr,e psfig, slashed}
\usepackage[normalem]{ulem}
\usepackage{tikzsymbols}
\usepackage{natbib}
\usepackage{float}
\usepackage{adjustbox}
\usepackage[compat=1.1.0]{tikz-feynman}
\usepackage{tikzsymbols}
\usepackage{tikz}
\usepackage{orcidlink}
\usepackage{bm}
\usepackage{latexsym}
\usepackage{subfig}
\usepackage{amsfonts}
\usepackage{bigints}
\usepackage{braket}
\usepackage{indent first}
\usepackage{multirow}
\usepackage{blindtext}
\usepackage{titlesec}
\useunder{\uline}{\ul}{}
 \usepackage{environ}
\usepackage{mathtools}
\usepackage{relsize}

\newcommand{\Trh}{T_\text{rh}}
\newcommand{\Tmax}{T_\text{max}}
\newcommand{\mdm}{m_{\rm DM}}
\newcommand{\gs}{g_\star}
\newcommand{\gss}{g_{\star s}}
\newcommand{\lmix}{\lambda_{\rm mix}}

\newcommand{\arh}{a_{\rm rh}}
\newcommand{\aend}{a_{\rm end}}
\newcommand{\Ndm}{N_{\rm dm}}
\newcommand{\ndm}{n_{\rm dm}}
\newcommand{\br}{\mathcal{B}}

\newcommand{\mphi}{m_\phi}
\newcommand{\st}{\sin\theta}
\newcommand{\stsq}{\sin^2 \theta}

\newcommand{\Br}{\operatorname{Br}}
\newcommand{\ctau}{c\tau_\phi}

\begin{document}
\title{Status of light inflaton: from inflation to laboratory}
\author[a]{Niloy Mondal\,\orcidlink{0009-0006-5837-9772},}
\author[b]{Shashwat Sharma\,\orcidlink{0009-0002-2266-0467},}
\author[c]{Mathew Thomas Arun\,\orcidlink{0000-0003-3264-3628},}
\author[b]{and Basabendu Barman\,\orcidlink{0000-0003-0374-7655}}
\affiliation[a]{\,\,Department of Physics, Indian Institute of Technology Guwahati, North Guwahati, 781039, India}
\affiliation[b]{\,\,Department of Physics, Indian Institute of Technology Hyderabad, Kandi, Sangareddy, Telangana-502285, India}
\affiliation[c]{\,\,School of Physics, Indian Institute of Science Education and Research, Thiruvananthapuram 695551,
Kerala, India}
\affiliation[d]{\,\,Department of Physics, School of Engineering and Sciences, SRM University-AP, Amaravati 522240, India}
\emailAdd{niloy18@iitg.ac.in}
\emailAdd{ph23resch11016@iith.ac.in}
\emailAdd{mathewthomas@iisertvm.ac.in}
\emailAdd{basabendu.b@srmap.edu.in}
\abstract{We investigate the viability of the light inflaton scenario in light of the latest inflationary constraints from the Atacama Cosmology Telescope (ACT), together with bounds from collider and intensity-frontier experiments searching for a feebly coupled light scalar with a sub-GeV mass. Assuming a quartic inflaton potential, we identify the region of parameter space consistent with the ACT observations and derive constraints on the inflaton mass and inflaton-Higgs mixing using results from NA62, KOTO, BaBar, Belle, LHCb, MATHUSLA, FASER2, SHiP, and neutral meson oscillations. We also explore the prospects for dark matter production during reheating within this framework, while remaining consistent with the inflationary observables.
}
\maketitle
\section{Introduction}
\label{sec:intro}
Cosmic inflation provides a compelling framework for understanding the earliest moments of the Universe and naturally resolves several shortcomings of the standard Big Bang picture, such as the horizon and flatness problems~\cite{Guth:1980zm,Linde:1981mu}. In its simplest realization, inflation is driven by a single scalar field---the inflaton---slowly rolling along an approximately flat potential and dominating the energy density of the Universe. The flatness of the potential is quantified through the slow-roll parameters, which are directly related to cosmological observables measured in the cosmic microwave background (CMB). Another key prediction of inflation is the generation of primordial gravitational waves (see, for example, Ref.~\cite{Caprini:2018mtu} for a review). Moreover, it has recently been shown that high-energy collider experiments may indirectly probe the inflationary sector through searches for new physics~\cite{Barman:2024nhr,Barman:2024tjt,Pradhan:2026maz}. A common assumption is that the inflaton is completely decoupled from the Standard Model (SM) at energies far below the inflationary scale, making direct laboratory tests of inflation impossible. However, the seminal ``Light Inflaton Hunter's Guide''~\cite{Bezrukov:2009yw} (based on the $\nu$MSM model introduced in~\cite{Shaposhnikov:2006xi}) systematically demonstrated that low-energy particle physics experiments can directly probe the inflaton sector within a simple chaotic inflation framework, through Higgs--inflaton mixing\footnote{Similar set-up has also been utilized to explore primordial GW and DM~\cite{Bezrukov:2014nza} or signatures of heavier inflaton (above $\sim$ 250 GeV) along with DM~\cite{Bezrukov:2020jmo}.}. Since then, this idea has been explored in several scenarios, including low-scale quartic hilltop inflation with a curvaton field~\cite{Bramante:2016yju}, scale-invariant gauged $U(1)_X$ extensions of the SM~\cite{Okada:2019opp}, and models where Higgs--inflaton mixing improves the stability of the electroweak vacuum~\cite{Ema:2017ckf}. 

Since inflation rapidly dilutes any pre-existing matter and radiation, the Universe emerges from the inflationary epoch in a highly diluted state, with its energy density stored predominantly in the nearly homogeneous inflaton condensate. To connect inflation with the standard hot Big Bang evolution, the inflaton must transfer its energy to the particle sector through a reheating phase, thereby restoring a thermal bath of relativistic particles and initiating the radiation-dominated era required for successful Big Bang nucleosynthesis (BBN). Beyond establishing the thermal history of the Universe, reheating plays a central role in determining the origin of its particle content. In particular, it can govern the production of dark matter (DM). A coherent picture of inflation and post-inflationary reheating therefore should address not only the production of all relevant degrees of freedom of the SM, but also cosmic origin of the beyond the SM entities.

Motivated by these, in this work we re-examine the viability of light inflaton, typically with mass $\lesssim 10$ GeV, in light of the latest cosmological observations from the Atacama Cosmology Telescope (ACT). The recent ACT Data Release 6 (DR6)~\cite{AtacamaCosmologyTelescope:2025blo,AtacamaCosmologyTelescope:2025nti} has significantly improved the precision of measurements of the high-$\ell$ CMB temperature and polarization power spectra. A combined analysis of ACT and Planck data (P-ACT) yields a scalar spectral index of $n_s = 0.9709 \pm 0.0038$, while the inclusion of CMB lensing and DESI baryon acoustic oscillation (BAO) data (P-ACT-LB) further tightens the constraint to $n_s = 0.9743 \pm 0.0034$. Compared to the Planck-only determination, these results indicate a noticeable upward shift in the preferred value of $n_s$, corresponding to a discrepancy at approximately the $2\sigma$ level. Given that the scalar spectral index serves as one of the primary discriminants among inflationary models, the new ACT measurements provide a timely opportunity to reassess the status of the light inflaton paradigm. On the other hand, Higgs mixed light inflaton can be effectively searched for at low-energy particle physics experiments, such as BaBar~\cite{BaBar:2001yhh,BaBar:2010oqg}, Belle~\cite{BelleII:2018jsg,Belle-II:2023esi}, LHCb~\cite{LHCb:2008vvz,LHCb:2016awg}, NA62~\cite{NA62:2017rwk,NA62:2021dod}, E787~\cite{Atiya:1992vh}, MATHUSLA~\cite{Curtin:2018mvb,MATHUSLA:2018bqv}, FASER/FASER2~\cite{FASER:2018eoc,Feng:2022inv,FPFWorkingGroups:2025rsc},  SHiP~\cite{SHiP:2020vbd,SHiP:2021nfo}. They provide powerful probes of light, feebly interacting particles beyond the SM. While NA62, BaBar/Belle and LHCb have much smaller detector lengths and are capable of searching for prompt decay of the inflaton, MATHUSALA, FASER2 and SHiP search for displaced vertices. Moreover, beyond on-shell production of the inflaton, FCNC effects at 1-loop, like $B^0$--$\bar{B}^0$ oscillations, through the $|\Delta F|=2$ effective operator, also constraint the inflaton parameter space. These neutral meson oscillation processes are not limited by the kinematic constraints that arise in the above mentioned prompt or displaced decay experiments. The primary objective of this work is therefore to investigate the potential of current and future intensity-frontier experiments to detect signatures of light inflaton, while ensuring consistency with the latest cosmological observations and inflationary constraints. We further discuss implications of particle DM production within the present framework. 

The remainder of this paper is organized as follows. In Sec.~\ref{sec:framework}, we describe the underlying model. Sec.~\ref{sec:inflation} is devoted to the inflationary predictions in light of ACT observations and the subsequent reheating dynamics. In Sec.~\ref{sec:bounds}, we discuss the laboratory constraints on the light inflaton scenario. Dark matter production within this framework is investigated in Sec.~\ref{sec:dm}. Finally, we conclude in Sec.~\ref{sec:concl}.
\section{The framework}
\label{sec:framework}
In order to introduce the inflaton, we extend the SM particle content with a SM gauge singlet scalar $\phi$. We consider the following action in the Jordan frame of reference~\cite{Shaposhnikov:2006xi,Bezrukov:2009yw,Bezrukov:2020jmo}, 
\begin{align}
&\mathcal{S}^{^{\mathcal{J}}}\supset\int \sqrt{-g}\,d^4x\,\left[\mathcal{L}_\phi+\mathcal{L}_{\rm int}+\mathcal{L}_{\rm grav}+\mathcal{L}_{\rm SM}\right]\,,
\label{eq:Action}
\end{align}
where we consider a FRW-like metric with mostly minus signature as the background. Here,
\begin{align}
& \mathcal{L}_{\phi}=\frac{1}{2}\,\partial_\mu\phi\,\partial^\mu\phi+\frac{1}{2}\,m_\phi^2\,\phi^2-\frac{\lambda_\phi}{4}\,\phi^4\,,
\label{eq:Lphi}
\\&
\mathcal{L}_{\rm int}=- \lambda_H\,\left(
H^\dagger H - \frac{\lambda_{\rm mix}}{\lambda_H}\,\phi^2
\right)^2\,,
\label{eq:Lint}
\\&
\mathcal{L}_{\rm grav}=
- \frac{M_P^2 + \xi_\phi\,\phi^2}{2}\,\mathcal{R}\,,
\label{eq:LSM}
\\&
\mathcal{L}_{\rm SM}=\Big|D_{\mu} H\Big|^2+\mathcal{L}_{\rm yuk}+i\bar{f}\slashed{\partial}\,f-\frac14\, g^{\mu\nu}\, g^{\lambda\rho}\, \mathcal{V}_{\mu\lambda}^{(a)}\, \mathcal{V}^{(a)}_{\nu\rho}\,,
\label{eq:Lgrav}
\end{align}
with $H=(1/\sqrt{2})\,\left(0~~~~h\right)^T$, in the unitary gauge, $\mathcal{V}^{(a)}$ denotes the SM gauge bosons (Abelian and non-Abelian) and $f$ stands for all SM fermions (quarks and leptons). The covariant derivative is defined as $D_\mu \equiv \partial_\mu - i\, g_2\, \tau^a\, W_\mu^a - i\, (g_1/2)\, Y\, B_\mu$, where $W_\mu$ and $B_\mu$ are the $SU(2)_L$ and $U(1)_Y$ gauge bosons, respectively, with corresponding $g_2$ and $g_1$ gauge coupling strengths, $Y$ is the hypercharge and $\tau^a = \sigma^a/2$ are the Pauli matrices. The SM Yukawa interactions are encoded in $\mathcal{L}_{\rm yuk}$. It is important to emphasize that, all fields are defined in the Jordan frame. The inflaton Lagrangian is given by Eq.~\eqref{eq:Lphi}, where we have considered a quartic potential for inflation. We assumed that the only source of scale symmetry violation is due to the negative mass term in the inflaton sector. Then, the negative quartic inflaton-Higgs coupling allows for the transfer of symmetry breaking into the SM sector. The interaction term corresponding to Eq.~\eqref{eq:Lint} gives rise to to spontaneous breaking of electroweak (EW) symmetry, and the SM Higgs field gains non-zero vacuum expectation value (VEV). During inflation the field $\phi$ is displaced from $\langle\phi\rangle$, and $\phi$ only acts as a classical background. The inflaton oscillates around the minimum of the potential during reheating, and eventually settles at $\phi=\langle\phi\rangle$.

To obtain the scalar spectrum, we minimize the scalar potential obtaining the following VEV condition,
\begin{align}\label{eq:vev}
& v_\phi=v_h\,\sqrt{\frac{\lambda_H}{\lambda_{\rm mix}}}\,, 
\end{align}
where $v_h=246$ GeV is the Higgs VEV. Using Eq.~\eqref{eq:vev}, one obtains the mass matrix
\begin{align}
&\mathcal{M}^2\simeq
\begin{pmatrix}
2\,\lambda_H\,v_h^{2} & -\,2\,\lambda_{\rm mix}\,v_h\,v_\phi \\[10pt]
-\,2\,\lambda_{\rm mix}\,v_h\,v_\phi & 2\,\lambda_\phi\,v_\phi^2
\end{pmatrix}\,,
\end{align}
in the small mixing approximation: $\lambda_{\rm mix}\ll\lambda_H$. This mass matrix can be diagonalized following,
\begin{align}
& 
\begin{pmatrix}
m_1^2 & 0 \\[10pt]    
0 & m_2^2
\end{pmatrix}
= R(\theta)^T\,\mathcal{M}^2\,R(\theta)\,,
\end{align}
where $$R(\theta)=
\begin{pmatrix}
\cos\theta & \sin\theta \\[10pt]    
-\sin\theta & \cos\theta
\end{pmatrix}$$ is the rotation matrix that rotates the weak eigenstates into the mass eigenstates through a rotation angle,
\begin{align}
& \tan2\theta=\frac{2\,v_h\,v_\phi\,\lmix}{v_\phi^2\,\lambda_\phi-v_h^2\,\lambda_H}\,.    
\end{align}
In the small mixing limit we have,
\begin{align}
& m_\phi=m_h\,\sqrt{\frac{\lambda_\phi}{\lmix}}\,,    
\end{align}
together with
\begin{align}
& m_\phi\simeq\frac{m_h}{\theta}\,\sqrt{\frac{\lmix}{\lambda_h}}=\frac{v_h}{\theta}\,\sqrt{2\,\lmix}\,,  
\end{align}
where
\begin{align}
 & \theta\simeq\frac{v_h}{v_\phi}\,\frac{\lmix}{\lambda_\phi}\,,  
\end{align}
considering $\lmix\ll\lambda_\phi$. The mass eigenstates turn out to be
\begin{align}\label{eq:massmatrix}
& 
\begin{pmatrix}
h_1 \\[10pt] h_2    
\end{pmatrix}
=\begin{pmatrix}
\cos\theta & -\sin\theta 
\\[10pt]
\sin\theta & \cos\theta
\end{pmatrix}
\begin{pmatrix}
h \\[10pt] \phi    
\end{pmatrix}\,.
\end{align}
Before moving on let us note that our analysis is performed at zero temperature, the scalar spectrum and inflaton decay widths are evaluated around the zero-temperature vacuum. A treatment of thermal corrections to the Higgs potential and the electroweak phase transition is beyond the scope of the present work. It is also worth mentioning that the reheating temperature may lie either above or below the electroweak (EW) scale, i.e., the Universe can still be undergoing reheating while the electroweak phase transition is already over, provided $\Tmax>T_{\rm EW}>\Trh$, where $T_{\rm EW}\simeq 160$ GeV is the temperature corresponding to the EW phase transition. 
\section{Inflationary predictions}
\label{sec:inflation}
We focus on a 1-loop effective quartic scalar potential for the inflaton,
\begin{align}
V(\phi)=\frac{1}{4}\,\lambda_\phi(\phi)\,\phi^4\,,    
\end{align}
where the running of $\lambda_\phi$ is described by its beta function
\begin{align}\label{eq:beta-lam}
& \beta(\mu)=\frac{d\lambda_\phi}{d\log\mu}\,,    
\end{align}
where $\mu$ is the renormalization scale. We can solve Eq.~\eqref{eq:beta-lam} as,
\begin{align}
&\lambda_\phi (\phi)=\lambda_\phi(\mu_0)+\sum_{k=1}^\infty\,\frac{\beta'^{(k)}}{k!}\,\log^k\left(\frac{\phi}{\mu_0}\right)\,    
\end{align}
where $\lambda_\phi(\mu_0)$ is the value of $\lambda_\phi$ at a scale $\mu_0$ and $\beta'^{(k)}$ represents the $k^{\rm th}$ derivatives of the beta function evaluated at the scale $\mu_0$. Keeping only the first order correction\footnote{As pointed out in~\cite{Marzola:2016xgb}, the quantum corrections to $\xi_\phi$ are negligible as long as $\xi_\phi<e^8$.},
\begin{align}
&\lambda_\phi(\phi)=\lambda_\phi(M_P)\,\left[1+\Delta_L(M_P)\,\ln\left(\frac{\phi}{M_P}\right)\right]\,.     
\end{align}
We fix the reference scale $\mu_0=M_P$. For brevity, here onward, we will refer $\lambda_\phi(M_P)$ as $\lambda_\phi$. The relative loop correction $\Delta_L\equiv\beta/\lambda_\phi$, is regarded as a free parameter\footnote{See, for example, Refs.~\cite{Allison:2014hna,Allison:2014zya,Ballesteros:2015noa,Marzola:2016xgb,Oda:2017zul,Racioppi:2018zoy,Bostan:2019fvk,Gialamas:2025kef,Wolf:2025ecy} for possible field content beyond the SM that could give rise to such radiative corrections.}. 

In order to derive bounds on the parameters of the scalar potential from CMB, we first perform a conformal (Weyl) transformation,
\begin{align}
& g_{\mu\nu}^{\mathbb{E}}=\omega^2\,g_{\mu\nu}^{\mathcal{J}}\,, & \omega(\phi)^2=1+\frac{\xi_\phi\,|\phi|^2}{M_P^2}\,.
\end{align}
Corresponding action for the inflaton in the Einstein frame can be written as\footnote{In the Einstein frame, the Weyl transformation induces direct interactions between the inflaton and the SM fields through Planck-suppressed operators. These interactions allow the inflaton to decay into kinematically accessible SM particles. However, the corresponding decay widths are suppressed by powers of the Planck scale and are typically negligible compared with the decays mediated by inflaton--Higgs mixing.},
\begin{align}
    \mathcal{S}^{^{\mathbb{E}}}=
\int \sqrt{-g^{\mathbb{E}}}\,d^4x\,\left[-\frac{M_P^2}{2}\,\mathcal{R}^{\mathbb{E}}+\frac{1}{2}\,g_{\mu\nu}^{\mathbb{E}}\,\partial^\mu\hat{\Phi}\,\partial^\nu\hat{\Phi}\,-V^{\mathbb{E}}(\hat{\Phi})\right]\,,
\label{eq:ActionE}
\end{align}
where one can calculate $\mathcal{R}^{\mathbb{E}}$ using the metric $g_{\mu\nu}^{\mathbb{E}}$ in the Einstein frame and the potential in the Einstein frame reads,
\begin{align}
    V^{\mathbb{E}}(\hat{\Phi})=\frac{V^{\mathcal{J}}(\phi(\hat{\Phi}))}{\omega^4(\hat{\Phi})}\,.
    \label{eq:erpot}
\end{align}
In the above equation we parametrized $\phi$ as\footnote{We work in the metric formulation as opposed to the Palatini
formulation, where the affine connection is varied independently of the
metric (see~\cite{Gialamas:2025kef} for ACT prediction on Palatini formulation).},
\begin{align}
    \frac{d\hat{\Phi}}{d\phi}\equiv \frac{1}{f(\phi)}\,\sqrt{f(\phi)+\frac{3}{2}M_P^2\,\left[f^{\prime}(\phi)\right]^2}=\frac{1}{\omega^2}\,\sqrt{\omega^2+6\,\xi_{\phi}^2\,\phi^2/M_P^2}\,,
    \label{eq:parametrize}
\end{align}
so that the kinetic energy in the Einstein frame can be made canonical with respect to the new field $\hat{\Phi}$, and $f(\phi)\equiv\omega^2$. The second term of Eq.~\eqref{eq:parametrize} appears due to the conformal transformation of the Ricci scalar. The solution of the differential equation (Eq.~\eqref{eq:parametrize}) with the condition $\hat{\Phi}(h=0)=0$  can be expressed as,
\begin{align}
\frac{\hat{\Phi}(\phi)}{M_P}
&= \frac{1}{\sqrt{\xi_\phi}}
\Bigg[\sqrt{1+6\,\xi_\phi}\,
\sinh^{-1}\left(\frac{\sqrt{\xi_\phi\,\left(1+6\,\xi_\phi\right)}\,\phi}{M_P}
\right)-\sqrt{6\,\xi_\phi}\,
\sinh^{-1}\left(
\frac{\sqrt{6}\,\xi_\phi\,\phi/M_P}
{\sqrt{1+\left(\frac{\sqrt{\xi_\phi}\,\phi}{M_P}\right)^2}}
\right)\Bigg]\,.
\end{align}
Since, at this stage, we are interested in the inflationary prediction for this framework, hence we concentrate only on the inflaton potential. Assuming $\xi_{\phi}\gg1$ and working in the inflationary regime characterized by large values $\phi\gg M_P/\xi_{\phi}$, from the above equation, the relation above can be inverted to express $\phi$ in terms of $\hat{\Phi}$. One then obtains,
\begin{align}
    \phi=\frac{M_P}{\sqrt{\xi_{\phi}}}\,\exp\left(\frac{\hat{\Phi}}{\sqrt{6}M_P}\right)\,,
\end{align}
where, for simplicity, we have made use of the identity $\sinh^{-1}{(x)}=\ln{\left[x+\sqrt{1+x^2}\right]}$. Using the above relation in Eq.~\eqref{eq:erpot}, we obtain, 
\begin{align}
    V^{\mathbb{E}}(\hat{\Phi})\simeq\frac{\lambda_{\phi}\,M_{P}^{4}}{4\xi^{2}_{\phi}}\,\left(1+e^{\frac{\hat{\Phi}}{\sqrt{6}M_P}}\right)^{-2}\,,
\end{align}
which is the expression for the scalar potential in the Einstein frame.
\subsection{Slow roll parameters and CMB bounds}
For any single field inflationary model, the slow-roll (SR) parameters can be defined as,
\begin{align}\label{eq:slowrollep}
&\epsilon_V\equiv \frac{1}{2} M_P^2 \left(\frac{(V^{\mathbb{E}})'}{V^{\mathbb{E}}}\right)^2\simeq\frac{8}{\left(\phi/M_P\right)^2\,\left[1+\xi_\phi\,(1+6\,\xi_\phi)\right]}
+\frac{4\Delta_L\,\left[1+\xi_\phi\,\left(\phi/M_P\right)^2\right]}{\left(\phi/M_P\right)^2\,\left[1+\xi_\phi\,(1+6\,\xi_\phi)\,\left(\phi/M_P\right)^2\right]}\,,
\\&
\eta_V\equiv M_P^2 \left(\frac{(V^{\mathbb{E}})''}{V^{\mathbb{E}}}\right)\simeq\frac{12+4\xi_\phi\,\left(\phi/M_P\right)^2\,\left[1-2\xi_\phi\,\left(\left(\phi/M_P\right)^2\,\left(1+6\,\xi_\phi\right)-6\right)\right]}{\left(\phi/M_P\right)^2\,\left[1+\xi_\phi\,(1+6\,\xi_\phi)\,\left(\phi/M_P\right)^2\right]^2}
\nonumber\\&
~~~~~~~~~~~~~~~~~~~~~~~~~~~~~~+\frac{\Delta_L\,\left[1+\xi_\phi\,\left(\phi/M_P\right)^2\right]\,\left[7+\xi_\phi\,\left(\phi/M_P\right)^2\,\left(36\,\xi_\phi+7\right)\right]}{\left(\phi/M_P\right)^2\,\left[1+\xi_\phi\,(1+6\,\xi_\phi)\,\left(\phi/M_P\right)^2\right]^2}\,,
\end{align}
where primes denote derivatives with respect to $\hat{\Phi}$ and we have kept up to the linear order of $\Delta_L$ for simplicity. Employing Eq.~\eqref{eq:slowrollep} together with the condition $\epsilon_V=1$ that indicates the end of inflation, we obtain,
\begin{align}\label{eq:phi-end}
\phi_{\rm end}\simeq M_P\,\left[\frac{4\Delta_L\,\xi_\phi+\sqrt{8\xi_\phi\,\left[2 \Delta_L^2\,\xi_\phi+(12\,  \xi_\phi+1)\,\Delta_L+4\,(6\,\xi_\phi +1)\right]}-1}{2\xi_\phi\,\left(1+6\,\xi_\phi\right)}\right]^{1/2}\,,
\end{align}
which corresponds to the filed value at the end of inflation. The number of e-folds, $N_k$, between the horizon exit to the end of the inflation is defined as,
\begin{align}
    N_k(\phi_k)&=\frac{1}{\sqrt{2} M_P}\int_{\phi_{\rm end}}^{\phi_k}\frac{d\phi}{\sqrt{\epsilon_{V}(\phi)}}\left(\frac{d\hat{\Phi}}{d\phi}\right)\,.
    \label{eq:Nk}
 \end{align}
By solving Eq.~\eqref{eq:Nk} numerically, $\phi_k$ can be obtained as a function of $N_k$ and $\xi_\phi$. In the large field limit $\sqrt{\xi_\phi}\,\phi/M_P\gg 1$, it is possible to obtain a simplified expression as,
\begin{align}\label{eq:Nk-large}
& N_k\simeq \frac{3\,\phi_k^2\,\xi_\phi}{4\,M_P^2}\,\left[1-\frac{\xi_\phi}{8}\,\frac{\phi_k^2}{M_P^2}\,\Delta_L\right]\,,    
\end{align}
considering $\phi_k\gg\phi_{\rm end}$, and $\epsilon_V\simeq \left[\frac{4}{3\,\xi_\phi^2}\,\left(\frac{M_P}{\phi_k}\right)^4\right]\,\left[1+\xi_\phi\,\Delta_L\,\left(\frac{\phi_k}{M_P}\right)^2\right]$. Under the slow-roll approximation, the key CMB observables, namely, the scalar spectral index ($n_s$), the tensor-to-scalar ratio ($r$), and the scalar amplitude ($A_s$), can be expressed in terms of the SR parameters as,
\begin{align}
&n_s=1-6\,\epsilon_V+2\,\eta_V\,,\nonumber\\&
r=16\,\epsilon_V\,.
\label{eq:nsr}
\end{align}
The amplitude of scalar power spectrum is given by,
\begin{align}\label{eq:As}
& A_s = \frac{1}{24\pi^2\,M_P^4}\,\frac{V^{\mathbb{E}}}{\epsilon_V}\Bigg|_{\phi_k}\,,    
\end{align}
which has a value of $\simeq 2.1\times 10^{-9}$~\cite{Planck:2018jri} measured at the pivot scale $k_\star\simeq 0.05\,\text{Mpc}^{-1}$. As a consistency check, in the large-field limit and for $\Delta_L \to 0$, Eqs.~\eqref{eq:slowrollep} and \eqref{eq:Nk} reduce to,
\begin{align}
& \epsilon_V \simeq \frac{3}{4N_k^2}\,, & \eta_V \simeq -\frac{1}{N_k}\,,
\end{align}
with
\begin{align}
\phi_k \simeq 2M_P\left(\frac{N_k}{3\xi_\phi}\right)^{1/2}\,.
\end{align}
These lead to
\begin{align}
& n_s\simeq 1-2/N_k\,, & r\simeq 12/N_k^2\,,
\end{align}
showing the characteristic attractor relations of non-minimally coupled quartic inflation. Further, in the large field limit,
\begin{align}
V^{\mathbb{E}}(\hat{\Phi})\simeq\frac{\lambda_{\phi}\,M_{P}^{4}}{4\xi^{2}_{\phi}}\,.   
\end{align}
One then obtains, from Eq.~\eqref{eq:As},
\begin{align}
& A_s\simeq\frac{1}{72\,\pi^2}\,\frac{\lambda_\phi\,N_k^2}{\xi_\phi^2}\implies\xi_\phi\simeq 4.6\times10^4\,\left(\frac{N_k}{55}\right)\,\lambda_\phi\,.    
\end{align}
\begin{figure}[htb!]
\centering
\includegraphics[scale=0.375]{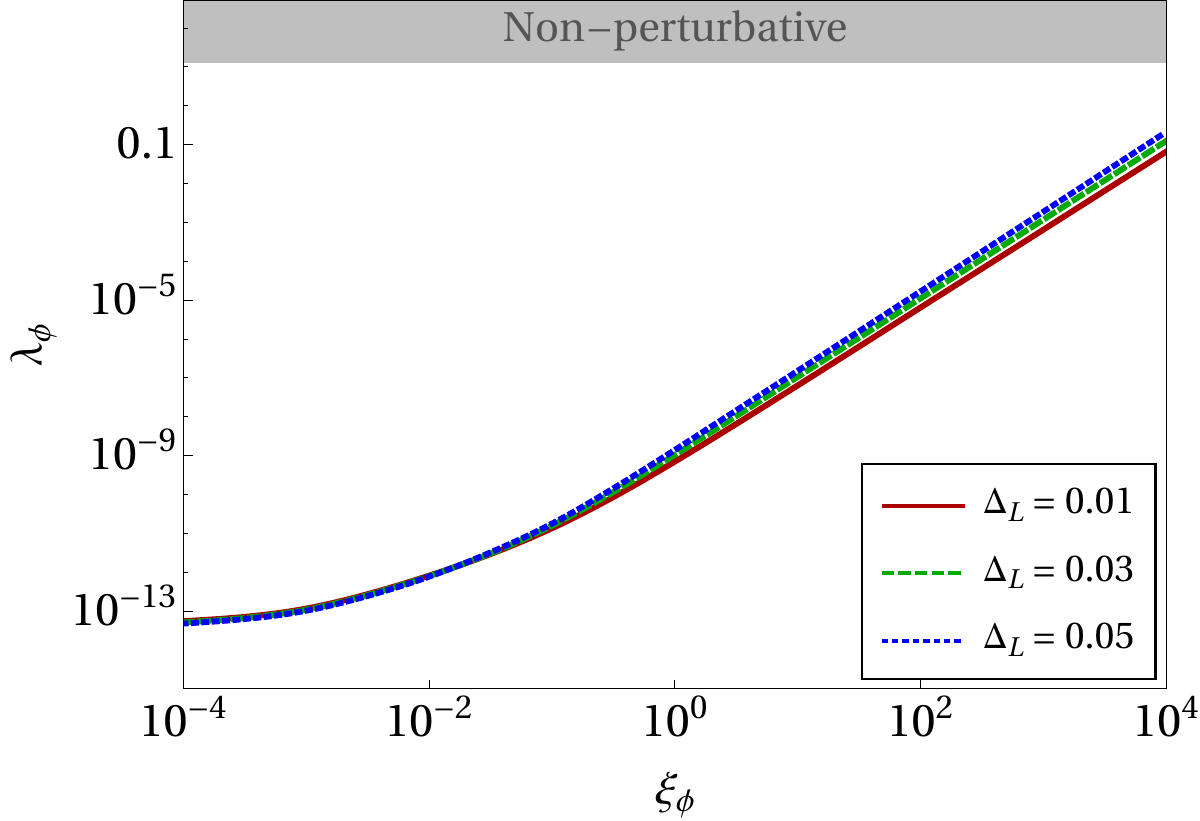}~\includegraphics[scale=0.375]{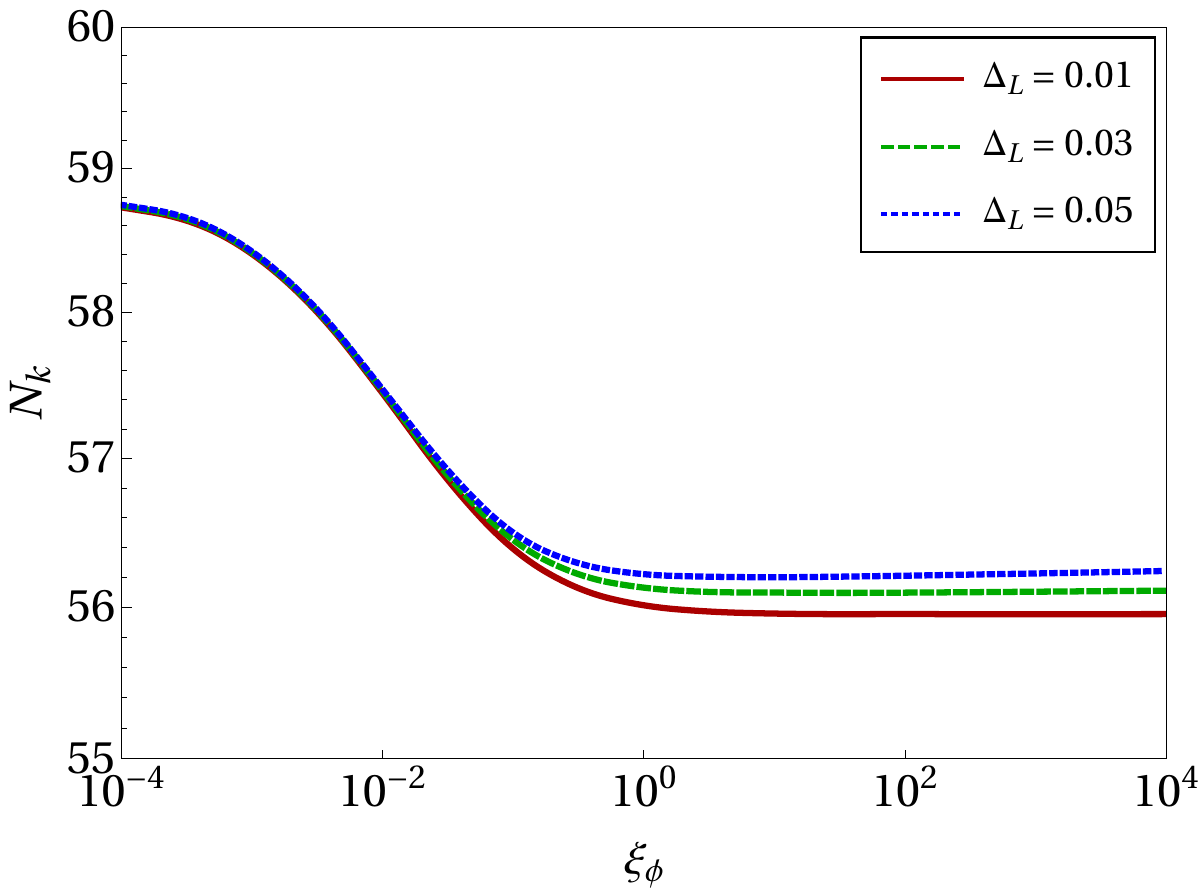}\\[10pt]
\includegraphics[scale=0.4]{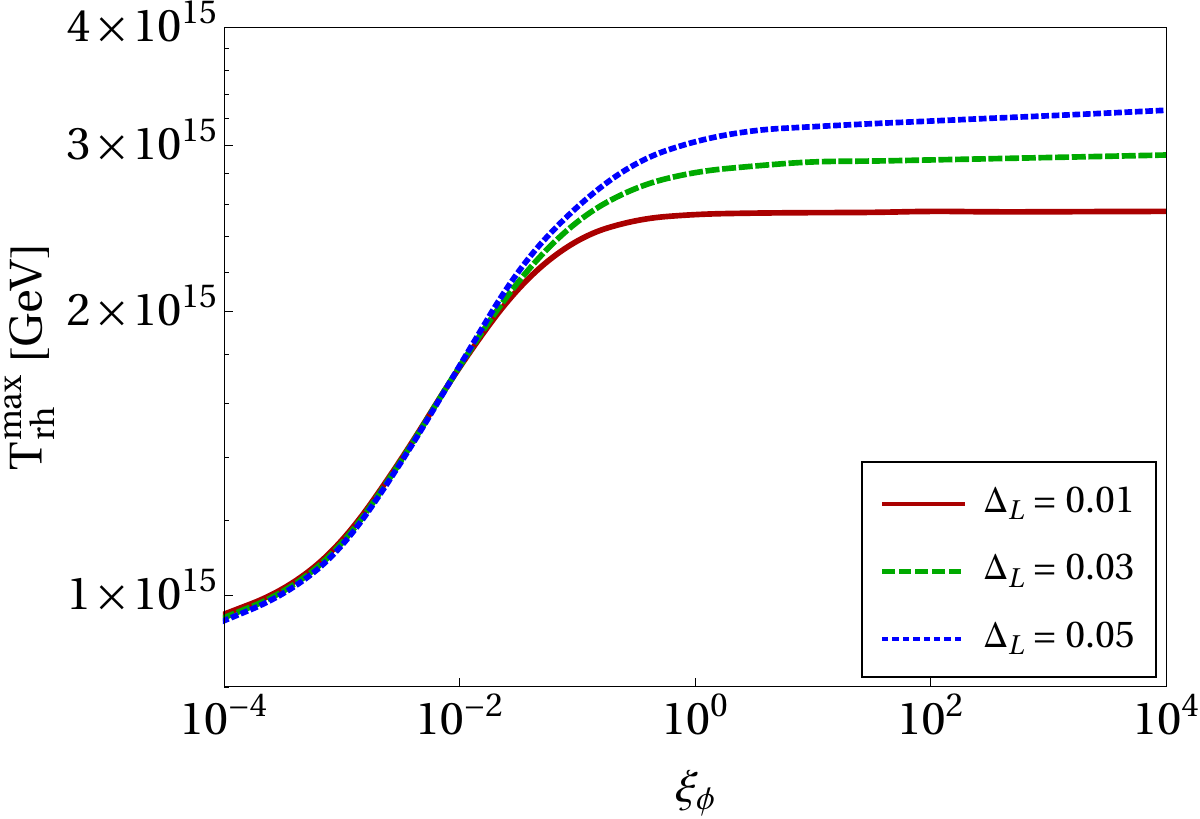}
\caption{Top left: Variation of the quartic coupling $\lambda_\phi$, as a function of the non-minimal coupling $\xi_\phi$, for different choices of the running parameter $\Delta_L$, shown via different patterns. Top right: number of e-folding (following Eq.~\eqref{eq:Nk}) as a function of $\xi_\phi$, for the same set of $\Delta_L$'s as in the left panel. Bottom: Maximum temperature during reheating, as a function of $\xi_\phi$ for different choices of the running parameter $\Delta_L$. In all cases we have chosen $n_s=0.965$, the central value provided by the ACT data.
}
\label{fig:xi}
\end{figure}
\begin{figure}[htb!]
\centering
\includegraphics[scale=0.65]{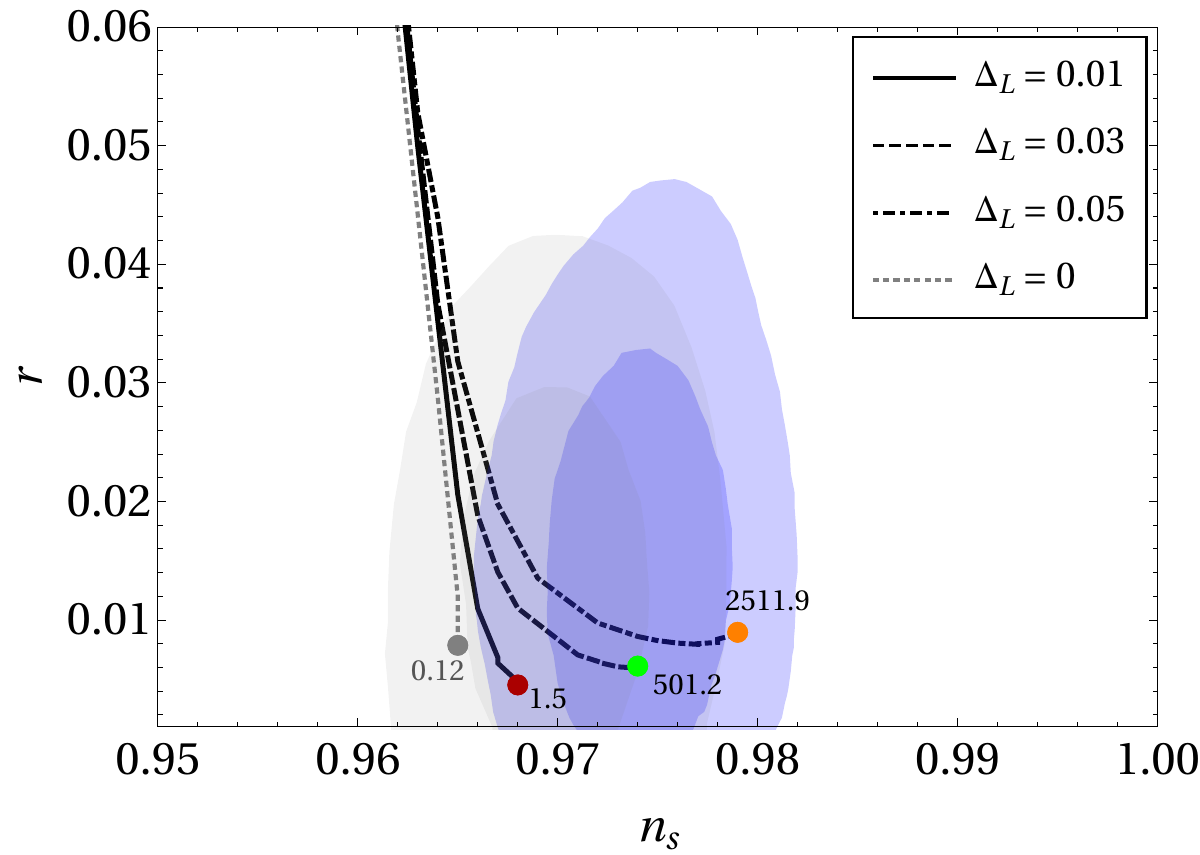}
\caption{Comparing model predictions with the combined constraints from Planck, ACT, DESI, and BICEP/Keck data in the bi-dimensional $[n_s-r]$ plane, for different choices of $\Delta_L$, as shown by different curves. The the value of the non-minimal coupling corresponding to each coloured points are mentioned. For all choices of $\xi_\phi$ and $\Delta_L$, we obtain $N_k=55$, and satisfy the minimum reheating temperature. The shaded regions in dark and light blue represent the 68\% $(1\sigma)$ and 95\% $(2\sigma)$ confidence levels, respectively. We also show Planck bounds in light gray.}
\label{fig:nsvr}
\end{figure}
The correlation between $\lambda_\phi$ and $\xi_\phi$ is shown in the {\it top left} panel of Fig.~\ref{fig:xi}, where we fix $n_s=0.965$, the central value reported by ACT. We observe that $\lambda_\phi$ does not vary significantly with $\Delta_L$, but increases steadily with $\xi_\phi$ in order to keep $n_s$ fixed. Over the range $\xi_\phi=\left[10^{-4}-10^4\right]$\footnote{The large values of the non-minimal coupling $\xi_\phi$ may raise potential concerns regarding perturbative unitarity. The validity of the effective description of inflation is commonly associated with a cutoff scale $\Lambda \sim M_P/\xi_\phi$ after inflation~\cite{Burgess:2009ea,Barbon:2009ya}, while during $\Lambda \sim M_P$. Recently, Ref.~\cite{Lebedev:2023zgw} demonstrated that, in the context of non-minimally coupled singlet-scalar inflation, perturbative unitarity may be violated for $\xi_\phi \gtrsim \mathcal{O}(10^2)$ once the production of inflaton quanta is taken into account, even in the absence of sizable interactions between the inflaton and other fields. In the parameter space favored by ACT data ($2\sigma$), we find, $\xi_\phi\sim\mathcal{O}(10^{-2})$ for $\Delta_L\gtrsim 0.01$.}, the perturbativity condition $\lambda_\phi<4\pi$ remains satisfied. 

For the same range of $\xi_\phi$, the {\it top right} panel displays the corresponding number of e-folds evaluated at the central value of $n_s$. First, for $\xi_\phi<10^{-2}$, $\Delta_L$ has essentially no impact on the e-folding number. This occurs because the net loop-corrected contribution $\xi_\phi\times \Delta_L$ in Eq.~\eqref{eq:Nk-large} is subdominant for small $\xi_\phi$ at fixed $\Delta_L$, so the SR trajectory remains nearly unchanged and $\phi_k$ stays close to its tree-level value. As a result, the curves for different $\Delta_L$ overlap.  In the limit $\xi_\phi\to\infty$, the curves become approximately independent of $\xi_\phi$, because for fixed $n_s$ the quantity $\epsilon_V$ effectively depends only on $\Delta_L$. In summary, at large $\xi_\phi$ the loop correction becomes increasingly relevant, causing the predictions to depart from the Starobinsky attractor and move toward the linear limit; larger $\Delta_L$ leads to an earlier departure from the Starobinsky solution.
\subsection{Connecting CMB parameters to reheating dynamics}
Considering inflaton decay exclusively into radiation, the coupled Boltzmann equations governing the inflaton and radiation energy densities are
\begin{align}
\label{eq:rhophi_rhord}
& \dot\rho_\phi + 3\,(1+w_{\rm rh})\,\mathcal{H}\,\rho_\phi
= -(1+w_{\rm rh})\,\Gamma_\phi\,\rho_\phi,
\nonumber\\
& \dot\rho_R + 4\,\mathcal{H}\,\rho_R
= +(1+w_{\rm rh})\,\Gamma_\phi\,\rho_\phi,	
\end{align}
with
\begin{equation}\label{eq:hubble}
\mathcal{H}^2=\frac{\rho_\phi+\rho_R}{3M_P^2}\,,
\end{equation}
where the dot indicates differentiation with respect to cosmic time $t$ and $w_{\rm rh}\equiv p_\phi/\rho_\phi$ is the effective EoS during reheating. At early times, the Hubble dilution term \mbox{$3(1+w_{\rm rh})H\rho_\phi$} dominates over the decay term, implying,
\begin{equation}\label{eq:rho-phi}
\rho_\phi(a)=\rho_{\rm end}\,\left(\frac{\aend}{a}\right)^{3\,(1+w_{\rm rh})}\,,
\end{equation}
where $a_{\rm end}$ denotes the scale factor at the end of inflation and 
\begin{align}
& \rho_{\rm end}\simeq V(\phi_{\rm end})=\frac{3\pi^2}{2}\,r\,A_s\,M_P^4\,, 
\end{align}
is the inflaton energy density at the end of inflation. Reheating ends when $\rho_\phi\simeq\rho_R$ at $a=a_{\rm rh}$ (at temperature $T=\Trh$). Using $\rho_R=(\pi^2/30)\,\gs(\Trh)\,\Trh^4$, the reheating temperature follows as
\begin{equation}
\label{eq:trh_w_nrh}
\Trh \simeq 
\left(\frac{90\,M_P^2\,\mathcal{H}_{\rm end}^2}{\pi^2\,\gs(\Trh)}\right)^{1/4}
\exp \left[-\frac{3}{4}N_{\rm rh}(1+w_{\rm rh})\right],
\end{equation}
where $N_{\rm rh}$ is the number of e-folds elapsed during reheating, with $N_{\rm rh}=0$ corresponding to instantaneous reheating. The Hubble scale at the end of inflation reads,
\begin{align}
& \mathcal{H}_{\rm end}\simeq\sqrt{\frac{\rho_{\rm end}}{3\,M_P^2}}\lesssim 4.5\times 10^{13}\,\text{GeV}\,, 
\end{align}
utilizing the upper bound on $r_{0.05}\lesssim 0.036$~\cite{Planck:2018jri}. Assuming that comoving entropy remains conserved from the end of reheating to today, the reheating temperature can also be related to the present CMB temperature ($T_0 = 2.735\ {\rm K}$) through~\cite{Dai:2014jja,Cook:2015vqa}
\begin{equation}
\label{eq:trh_w_nrh_ne}
\Trh = 
\left(\frac{43}{11\,\gss(\Trh)}\right)^{1/3}
T_0\,\frac{\mathcal{H}_k}{k_\ast}\,
\exp[-(N_k+N_{\rm rh})]\,,
\end{equation}
while the maximum reheating temperature $\Trh^{\rm max}$ corresponds to $N_{\rm rh}\to0$, i.e., instant reheating,
\begin{align}\label{eq:Tmax}
& \Trh^{\rm max}\simeq\left[\frac{30\,\rho_{\rm end}}{\pi^2\,\gs(\Trh^{\rm max})}\right]^{1/4}\simeq 10^{15}\,\text{GeV}\,,    
\end{align}
considering $n_s=0.965$. We note, for a given $n_s$, the maximum temperature changes very slowly with $\xi_\phi$, as one can see from the bottom panel of Fig.~\ref{fig:xi}. This is expected from Eq.~\eqref{eq:Tmax}, which is only sensitive to the CMB observables. By equating Eqs.~\eqref{eq:trh_w_nrh} and~\eqref{eq:trh_w_nrh_ne}, one obtains a relation between $N_k$ and the reheating temperature as~\cite{Dai:2014jja,Cook:2015vqa},
\begin{align}
\label{eq:Ninf_reheating}
N_k =\,
\log\bigg[
\left(\frac{43}{11\,\gss(\Trh)}\right)^{ 1/3}
T_0\,\frac{\mathcal{H}_k}{k_\ast}\,
\Trh^{\frac{1-3\,w_{\rm rh}}{3\,(1+w_{\rm rh})}}
\left(\frac{\pi^2 \gs(\Trh)}{90\,M_P^2\,\mathcal{H}_{\rm end}^2}\right)^{ \frac{1}{3\,(1+w_{\rm rh})}}
\bigg].
\end{align}
Comparing Eq.~\eqref{eq:Ninf_reheating} with model-dependent expressions for $N_k$ allows one to construct a direct correspondence between the inflationary parameters and the post-inflationary expansion history. The physical content of Eq.~\eqref{eq:Ninf_reheating} is twofold: it encodes the properties of the inflationary potential (together with CMB observables) through $N_k$, while simultaneously incorporating particle physics effects via the reheating temperature $\Trh$, which is set by the inflaton decay channels encoded in $\Gamma_\phi$. Consequently, the decay modes via scalar mixing become dominant. 
\begin{figure}
\centering
\includegraphics[scale=0.48]{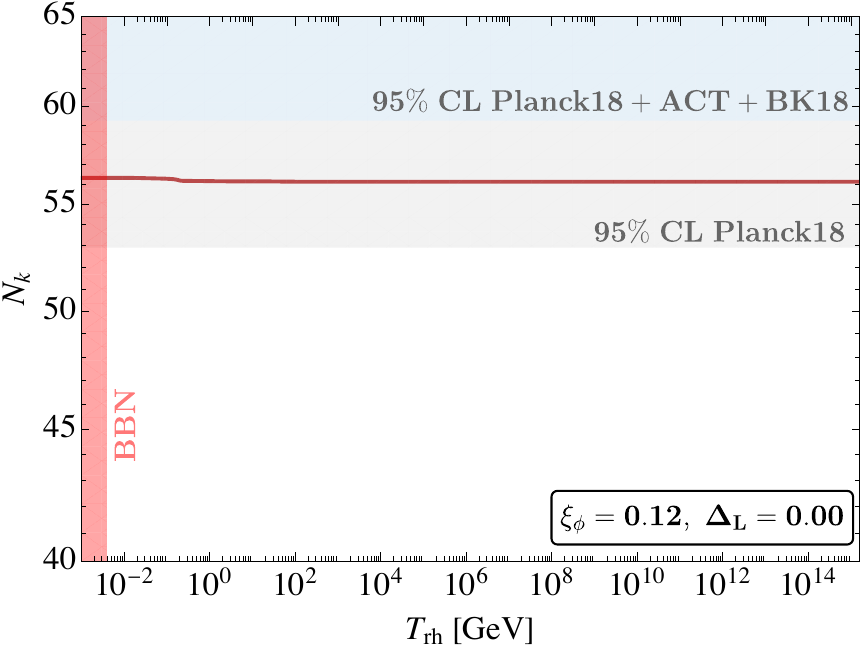}~\includegraphics[scale=0.48]{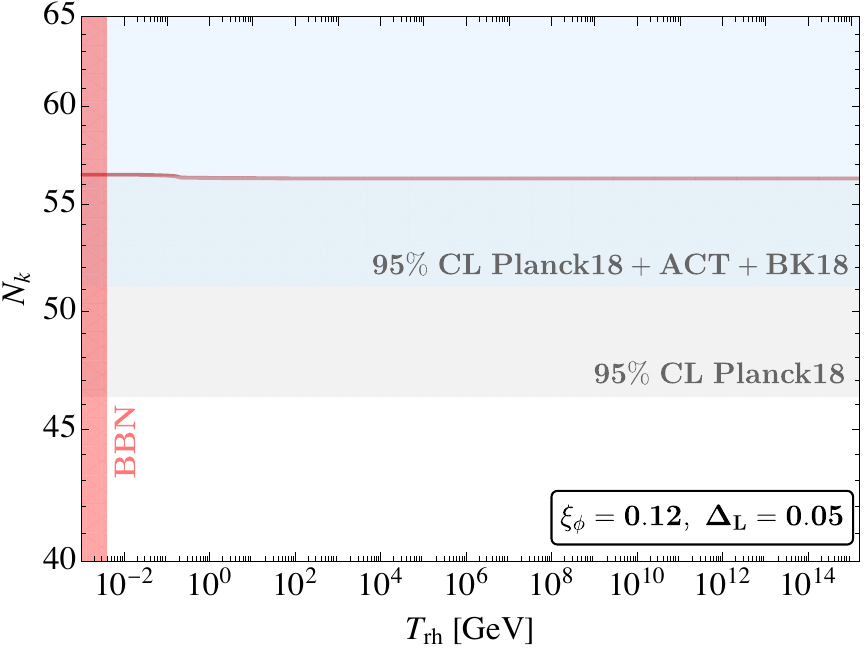}\\[10pt]
\includegraphics[scale=0.48]{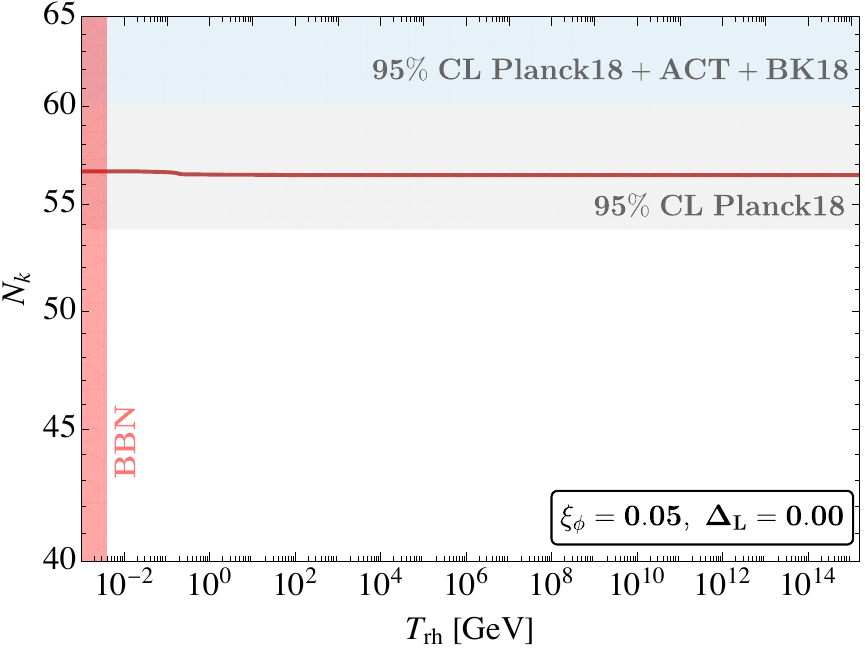}~\includegraphics[scale=0.48]{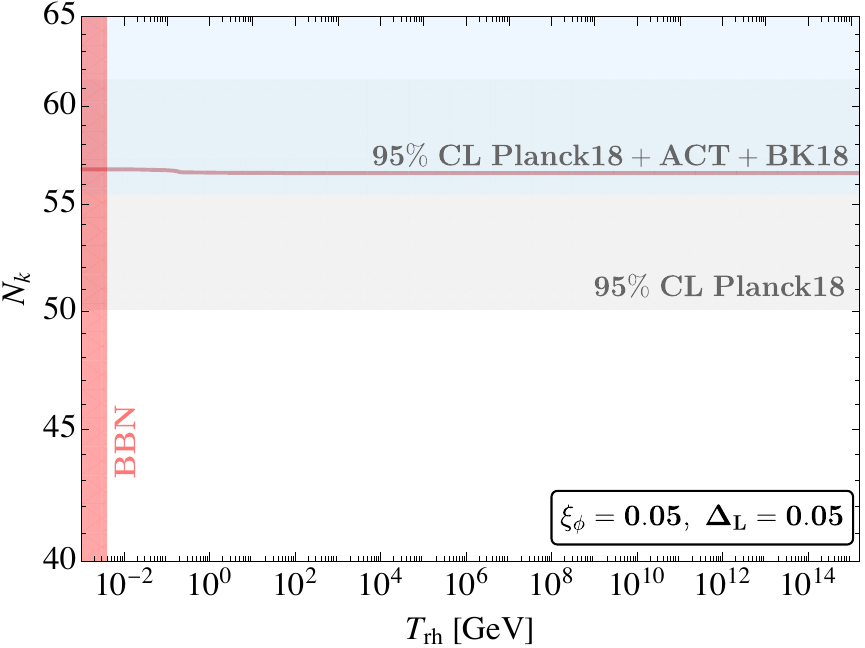}\\[10pt]
\includegraphics[scale=0.48]{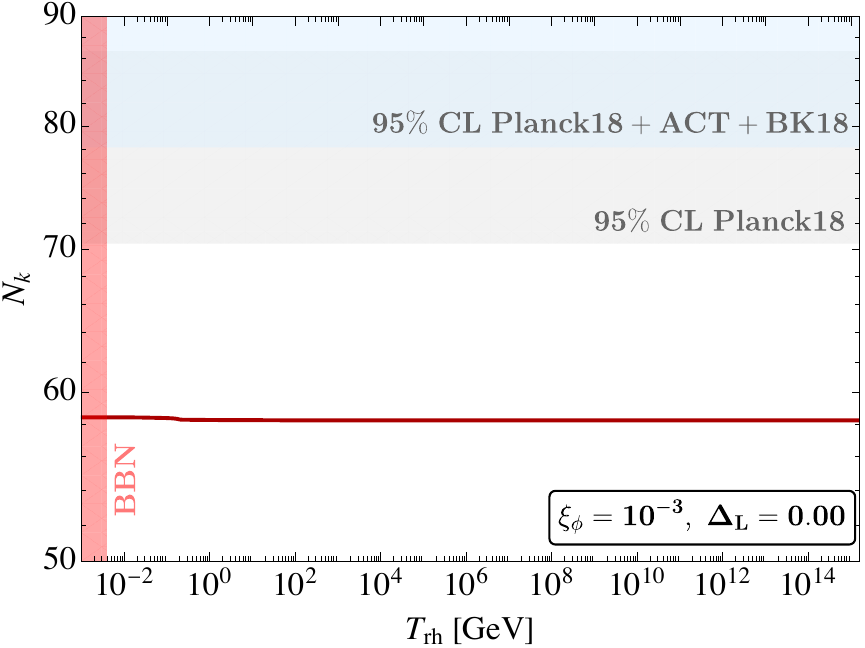}~\includegraphics[scale=0.48]{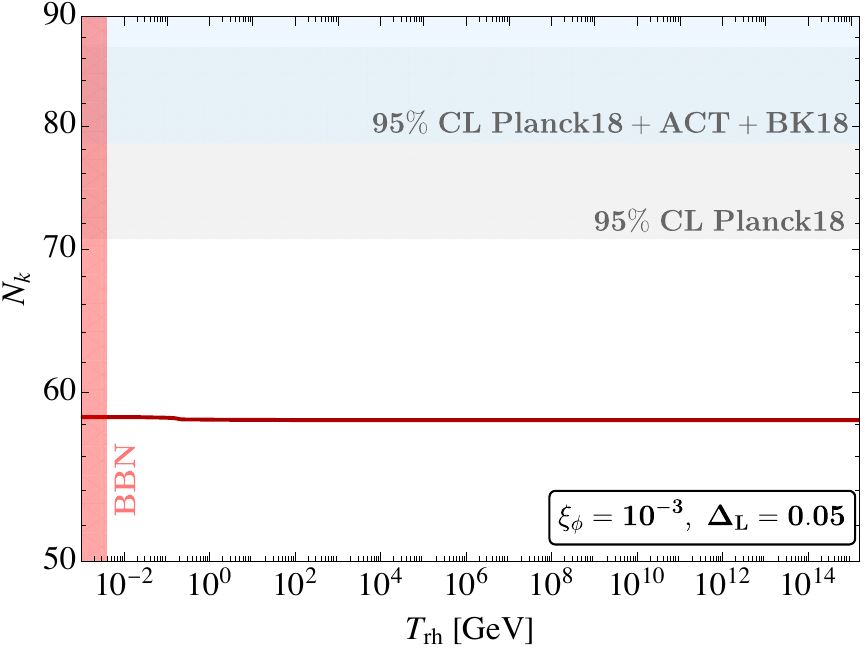}
\caption{Variation of the number of e-folds with the reheating temperature, computed using Eq.~\eqref{eq:Ninf_reheating}, for different values of the non-minimal coupling and loop-correction contributions for $w_{\rm rh}=1/3$. The red shaded region is excluded by the lower bound on the reheating temperature set by BBN ($T_{\rm RH}^{\rm min}=T_{\rm BBN}\simeq4 \mathrm{MeV}$). The blue and gray shaded regions denote the $2\sigma$ constraints from the combined Planck+ACT+DESI+BICEP/Keck data and from Planck data alone, respectively.}
\label{fig:Nk-Trh}
\end{figure}

In Fig.~\ref{fig:nsvr} we present the model predictions together with the ACT observational constraints. We fix $w_{\rm rh}=1/3$, effectively corresponding to the EoS parameter of standard radiation domination and scan over a range of non-minimal coupling $\xi_\phi\in[10^{-5}-10^4]$. As we can see, the predictions for $\Delta_L=0$ lie just outside the 2$\sigma$ contour of ACT data but within the 2$\sigma$ contour of Planck data, whereas larger values of $\Delta_L$ shift the results comfortably within the 1$\sigma$ region. In all cases, we find $N_k\simeq 55$ to be consistent with the ACT data. Although Fig.~\ref{fig:nsvr} corresponds to a reheating temperature $\Trh=4$ MeV, which sets the lower bound, however for $\Trh=\Tmax$ the outcome does not change. The justification for this statement can be understood from Fig.~\ref{fig:Nk-Trh}, where the number of e-folds is plotted as a function of the reheating temperature $T_{\rm rh}$ for $w_{\rm rh}=1/3$, using Eq.~\eqref{eq:Ninf_reheating}. With this EoS, the dependence of $N_{k}$ on $T_{\rm rh}$ arises only through the effective number of relativistic degrees of freedom. Consequently,$N_{k}$ is nearly independent of the reheating temperature, exhibiting only a mild variation around the QCD phase transition. The observational bounds on $N_{k}$ are obtained from Eq.~\eqref{eq:Nk}  where the field values $\phi_{k}$ and $\phi_{\rm end}$ are determined numerically by imposing $n_{s}=0.965$ (the central value from ACT data) and the end-of-inflation condition $\epsilon_{V}=1$, respectively. To maintain $n_{s}$ and $\epsilon_{V}$ within a given range, an increase in $\xi_{\phi}$ (and/or $\Delta_{L}$) must be compensated by smaller values of $\phi_{k}$ and $\phi_{\rm end}$. This in turn reduces $N_{k}$ (see Eq.~\eqref{eq:Nk}).  Consequently, as shown in Fig.~\ref{fig:Nk-Trh}, increasing $\xi_{\phi}$ while keeping $\Delta_{L}$ fixed shifts the allowed range of $N_{k}$ toward lower values. An analogous behavior is observed when $\Delta_{L}$ is increased for fixed values of $\xi_{\phi}$. Fig.~\ref{fig:Nk-Trh} shows that, even in the presence of a non-minimal coupling as large as $\xi_{\phi}\sim\mathcal{O}(10^{-1})$, a non-zero loop contribution is still required for the $N_{k}$ to lie within the $1\sigma$ region preferred by the combined Planck+ACT+DESI+BICEP/Keck data.

Now, substituting Eq.~\eqref{eq:rho-phi} into the second line of Eq.~\eqref{eq:rhophi_rhord} we obtain an approximate solution for the radiation energy density as,
\begin{align}
&\rho_R(a)\simeq \frac{2}{\sqrt{3}}\,\sqrt{\rho_{\rm end}\,M_P^2\,\Gamma_\phi^2(\theta)}\,\left[1-\left(\frac{\aend}{a}\right)^2\right]\,,    
\end{align}
where $\Gamma_\phi(\theta)$ is the inflaton decay rate, as a function of the scalar mixing angle, computed in the Einstein frame. It is important to note that we assume reheating proceeds through perturbative decays\footnote{Non-perturbative particle production (preheating) in this set-up has been discussed, for example, in~\cite{Bezrukov:2020jmo}.} of the inflaton mass eigenstate into SM particles via its Higgs mixing. Because of the non-minimal coupling the inflaton naturally couples to all the SM fields, however all such decay rates are suppressed by $1/M_P^2$. The reheating temperature then reads,
\begin{align}
& \Trh\equiv T(\arh)\simeq \left[\left(\frac{60\,M_P}{\pi^2\,\gs(\Trh)}\right)\,\Gamma_\phi(\theta)\,\sqrt{\frac{\rho_{\rm end}}{3}}\,\left(\frac{\arh}{\aend}\right)^2\right]^{1/4}\,.    
\end{align}
On utilizing Eq.~\eqref{eq:rho-phi},
\begin{align}\label{eq:treh-gamma}
&\Trh\simeq \left(\frac{40}{\pi^2\,\gs(\Trh)}\right)^{1/4}\,\sqrt{\Gamma_\phi(\theta)\,M_P}\,.    
\end{align}
Now, $\Trh$ is bounded: $4\,\text{MeV}\,(T_{\rm BBN})\lesssim\Trh\lesssim 10^{15}\,\text{GeV}\,(\Trh^{\rm max})$, which in turn provides a bound on the inflaton decay rate $1.04\times 10^{-23}\,\text{GeV}\lesssim\Gamma_\phi\lesssim 2.06\times 10^{12}\,\text{GeV}$, or, equivalently, on the inflaton lifetime $3.2\times 10^{-37}\,\text{s}\lesssim\tau_\phi\equiv1/\Gamma_\phi\lesssim 0.06\,\text{s}$. Note that, for $\Gamma_\phi\simeq 6.7\times 10^{-42}$ GeV, the inflaton lifetime becomes $\sim 10^{17}$ s, approximately comparable to the lifetime of the Universe. 
\section{Laboratory bounds on light inflaton}
\label{sec:bounds}
Following Eq.~\eqref{eq:massmatrix}, the light inflaton mixes with the Higgs boson and consequently acquires Higgs-like couplings to the SM fermions \(f\), suppressed by the scalar mixing angle \(\theta\),
\begin{equation}
\mathcal{L}_{\phi ff}\supset
-\sin\theta\,\frac{m_f}{v_h}\,\phi\,\bar f f\,.
\label{eq:yukawa}
\end{equation}
This same mixing angle governs both the production of $\phi$ in meson decays and its visible decays into SM particles. The total decay width can therefore be written as,
\begin{equation}
\Gamma_\phi(m_\phi,\theta)=
\sin^2\theta\,\Gamma_{h,\rm SM}(m_\phi)\,,
\label{eq:total_width}
\end{equation}
where $\Gamma_{h,\rm SM}(m_\phi)$ denotes the decay width of an off-shell SM Higgs boson evaluated at mass $m_\phi$. The corresponding proper decay length is then,
\begin{equation}
\ctau=\frac{\hbar c}
{\sin^2\theta\,\Gamma_{h,\rm SM}(m_\phi)},
\label{eq:ctau}
\end{equation}
with $\hbar c\simeq1.97\times10^{-16}\,{\rm GeV\,m}$. For a detector of length $L$ and an inflaton boost factor $\beta_\phi\gamma_\phi$, the transition between detector-stable and visibly-decaying behaviour occurs when the boosted decay length equals the detector size, $\beta_\phi\gamma_\phi\,\ctau=L$. This defines a critical mixing angle,
\begin{equation}
\sin^2\theta_{c}=\frac{\beta_\phi\gamma_\phi\,\hbar c}
{\Gamma_{h,\rm SM}(m_\phi)\,L}\,.
\label{eq:critical_mixing}
\end{equation}
For $\sin^2\theta \ll \sin^2\theta_{c}$, the inflaton typically escapes the detector before decaying and appears as missing energy. In contrast, for $\sin^2\theta \gg \sin^2\theta_{c}$, it decays promptly inside the detector. Using the relevant SM Higgs partial widths, the decay rates of the inflaton are,
\begin{align}
\Gamma_\phi(m_\phi,\theta)
=N_c\,\frac{m_\phi\,\sin^2\theta}{8\pi v_h^2}
\begin{dcases}
m_\ell^2\left(1-\frac{4m_\ell^2}{m_\phi^2}\right)^{3/2}\,,
& \text{for leptons}\,,
\\[10pt]
m_q^2\left(1-\frac{4m_q^2}{m_\phi^2}\right)^{3/2}\,,
& \text{for quarks}\,,
\end{dcases}
\end{align}
where $N_c=1(3)$ is the colour factor for leptons (quarks), $m_{\ell(q)}$ denotes the mass of the corresponding SM lepton (quark). For $m_\phi \lesssim 2\,{\rm GeV}$, the perturbative description becomes inadequate and must be supplemented by hadronic decay modes such as $\phi\to\pi\pi$, $\phi\to K\bar K$ and $\phi \to \eta\eta$, which are described using scalar hadronic form factors. Matching the partial widths onto chiral perturbation theory (ChPT), one obtains~\cite{Bramante:2016yju},
\begin{eqnarray}
  \Gamma(\phi\to\pi^+\pi^- + \pi^0\pi^0)
    &= \sin^2\!\theta_\phi\;\frac{3(m_u^2+m_d^2)\,m_\phi}{8\pi v^2}\,
       \beta_\pi^3,
       \label{eq:Gamma_pipi} \\[4pt]
  \Gamma(\phi\to K^+K^- + K^0\bar{K}^0)
    &= \sin^2\!\theta_\phi\;\frac{27}{13}\frac{m_s^2\,m_\phi}{8\pi v^2}\,
       \beta_K^3,
       \label{eq:Gamma_KK} \\[4pt]
  \Gamma(\phi\to\eta\eta)
    &= \sin^2\!\theta_\phi\;\frac{12}{13}\frac{m_s^2\,m_\phi}{8\pi v^2}\,
       \beta_\eta^3,
       \label{eq:Gamma_etaeta}
\end{eqnarray}
where $\beta_P=\sqrt{1 - 4m_P^2/m_\phi^2}$, and the numerical coefficients arise from the flavour-octet structure of the ChPT scalar currents\footnote{Note that, next-to leading order ChPT for $\phi \to \pi \pi, KK$ and $\eta \eta$~\cite{Gorbunov:2023lga} will improve our results moderately, we use the LO ChPT for simplicity.}. All three channels are kinematically forbidden below their respective two-body thresholds $m_\phi < 2m_P$.  

The production of $\phi$ from meson decays is induced by a one-loop Higgs-penguin diagram, which generates flavour-changing scalar transitions. For $B$ mesons, the relevant process is $b\to s\phi$, while for kaons it is $s\to d\phi$. The effective interaction responsible for the $b\to s\phi$ transition is
\begin{equation}
\mathcal{L}_{bs\phi}\supset C_{bs}\,\phi\,\bar s_L b_R
+ C_{bs}^*\,\phi\,\bar b_R s_L\,,
\label{eq:Lbs}
\end{equation}
where the Wilson coefficient obtained from matching the top--$W$ Higgs-penguin diagram is
\begin{equation}
C_{bs}=\st\,\frac{3g^2}{64\pi^2}\,\frac{m_b\,m_t^2}{m_W^2\,v_h}\,V_{ts}^*V_{tb}=\st\,\frac{3\,m_b\,m_t^2}
{16\pi^2\,v_h^3}\,V_{ts}^*V_{tb}\,.
\label{eq:Cbs}
\end{equation}
Neglecting the strange-quark mass, the corresponding partonic decay width is,
\begin{equation}
\Gamma(b\to s\phi)
\simeq
\frac{|C_{bs}|^2 m_b}{32\pi}
\left(1-\frac{\mphi^2}{m_b^2}\right)^2\,.
\label{eq:b_width}
\end{equation}
This leads to the standard estimate for the inclusive branching fraction,
\begin{equation}
\Br(B\to X_s\phi)\simeq
6.2\,\left(1-\frac{\mphi^2}{m_B^2}\right)^2
\stsq\,,
\label{eq:B_inclusive}
\end{equation}
which is valid for $\mphi\lesssim m_B$. For the exclusive decay channel $B\to K\phi$, the relevant scalar-current matrix element is
\begin{equation}
\langle K|\bar s b|B\rangle
=\frac{m_B^2-m_K^2}{m_b-m_s}\,
f_0^{BK}(\mphi^2)\,,
\label{eq:BK_formfactor}
\end{equation}
where $f_0^{BK}$ denotes the scalar form factor. The corresponding decay width then reads,
\begin{equation}
\Gamma(B\to K\phi)=\frac{|C_{bs}|^2}{64\pi m_B^3}\,\lambda^{1/2}(m_B^2,m_K^2,\mphi^2)
\left[\frac{m_B^2-m_K^2}{m_b-m_s}\,f_0^{BK}(\mphi^2)\right]^2\,,
\label{eq:B_exclusive_width}
\end{equation}
where the Källén function is defined as: $\lambda(a,b,c)=a^2+b^2+c^2-2ab-2ac-2bc$. The exclusive branching fraction is then given by,
\begin{align}
\Br(B\to K\phi)=\tau_B\,\Gamma(B\to K\phi)\, ,    
\end{align}
where $\tau_B$ is the decay time of the B meson.

Similarly, the effective interaction responsible for the flavor-changing transition $s\to d\phi$ is given by,
\begin{equation}
\mathcal{L}_{sd\phi}\supset C_{sd}\,\phi\,\bar d_L s_R
+ C_{sd}^*\,\phi\,\bar s_R d_L\,,
\label{eq:Lsd}
\end{equation}
where
\begin{equation}
C_{sd}=\st\,\frac{3g^2}{64\pi^2}
\frac{m_s\,m_t^2}{m_W^2 v_h}\,
V_{ts}^*V_{td}=\st\,\frac{3m_s m_t^2}{16\pi^2\,v_h^3}\,
V_{ts}^*V_{td}\,.
\label{eq:Csd}
\end{equation}
The hadronic matrix element relevant for the decay $K^+\to\pi^+\phi$ can be written as,
\begin{equation}
\langle\pi^+|\bar d s|K^+\rangle
=\frac{m_K^2-m_\pi^2}{m_s-m_d}\,
f_0^{K\pi}(\mphi^2)\,,
\label{eq:Kpi_formfactor}
\end{equation}
which leads to the decay width
\begin{equation}
\Gamma(K^+\to\pi^+\phi)=\frac{|C_{sd}|^2}{64\pi m_K^3}\,\lambda^{1/2}(m_K^2,m_\pi^2,\mphi^2)\,\left[
\frac{m_K^2-m_\pi^2}{m_s-m_d}
\,f_0^{K\pi}(\mphi^2)
\right]^2\,,
\label{eq:K_width}
\end{equation}
valid for $\mphi<m_K-m_\pi$. Substituting the numerical values of the CKM matrix elements and quark masses, the branching ratio can be expressed in the compact form,
\begin{equation}
\Br(K^+\to\pi^+\phi)\simeq
1.6\times10^{-3}\,\stsq\,\frac{\lambda^{1/2}(m_K^2,m_\pi^2,\mphi^2)}
{m_K^2-m_\pi^2}\,\left[\frac{f_0^{K\pi}(\mphi^2)}{f_0^{K\pi}(0)}\right]^2\,,
\label{eq:K_numeric}
\end{equation}
where $f_0^{K\pi}(0)\simeq 1$. For the neutral kaon channel, we can now parameterize the branching ratio as,
\begin{equation}
\Br(K_L\to\pi^0\phi)=\mathcal{B}_{K_L}(\mphi)\,\stsq\,,
\label{eq:KL_param}
\end{equation}
and derive bounds on the $\st$ using the experimental upper limit on the invisible decay mode $K_L\to\pi^0+{\rm inv.}$
\subsection{Decay probabilities and lifetime}
\label{sec:probability_regimes}
The same inflaton production mechanism can give rise to three different experimental signatures, depending on where the scalar decays: \emph{invisible}, \emph{prompt visible}, or \emph{displaced visible}. Consequently, the decay probability associated with each signature is just as important as the production branching fraction when translating experimental constraints into the $(m_\phi,\sin\theta)$ parameter space.

For a detector region extending over a distance $L\in \left[L_{\rm in},\,L_{\rm out}\right]$ along the scalar trajectory, the probability that $\phi$ decays inside the detector volume is given by~\cite{Curtin:2018mvb}
\begin{eqnarray}
P_{\rm dec}(L_{\rm in},L_{\rm out})
&=&
\exp\left[-\frac{L_{\rm in}}{\beta_\phi\gamma_\phi c\tau}\right]
-
\exp\left[-\frac{L_{\rm out}}{\beta_\phi\gamma_\phi c\tau}\right]
\nonumber\\
&\simeq&
\frac{L_{\rm out}-L_{\rm in}}
{\beta_\phi\gamma_\phi c\tau}\,,
\qquad
{\rm for}\;
(L_{\rm out}-L_{\rm in})
\ll
\beta_\phi\gamma_\phi c\tau\,.
\label{eq:Pdec}
\end{eqnarray}
Here, $\beta_\phi\gamma_\phi=|\vec p_\phi|/m_\phi$ is the boost factor of the scalar as defined previously. The first exponential represents the probability that $\phi$ survives up to the detector entrance, while the second gives the probability that it survives beyond the detector exit. Their difference therefore corresponds to the probability that the decay occurs inside the fiducial volume. This quantity is the relevant acceptance factor for displaced-vertex searches such as SHiP, MATHUSLA, FASER, and displaced searches at LHCb. If the scalar does not decay inside the detector, it may escape entirely. For a detector with effective length $L_{\rm det}$, the escape probability is
\begin{equation}
P_{\rm esc}(L_{\rm det})=\exp\left[-\frac{L_{\rm det}}{\beta_\phi\gamma_\phi c\tau}\right]\,.
\label{eq:Pesc}
\end{equation}
This factor becomes important when $\phi$ decays visibly in principle but is sufficiently long-lived to leave the detector before decaying. If the scalar also possesses a genuine invisible decay mode, the total probability for the event to appear invisible becomes,
\begin{equation}
P_{\rm inv}={\rm Br}(\phi\to{\rm inv.})+{\rm Br}(\phi\to{\rm vis.})\,P_{\rm esc}(L_{\rm det})\,.
\label{eq:Pinv}
\end{equation}
For a stable scalar, or for one that predominantly decays into invisible dark sector states, $P_{\rm inv}\simeq 1$. In contrast, for a Higgs-mixed scalar with only visible Standard Model decay channels, ${\rm Br}(\phi\to{\rm inv.})=0$, so that $P_{\rm inv}=P_{\rm esc}$.

Prompt-resonance searches require the scalar to decay very close to the production point. If $L_{\rm prompt}$ denotes the maximum displacement still reconstructed as prompt, the corresponding probability is
\begin{equation}
P_{\rm prompt}=1-\exp\left[-\frac{L_{\rm prompt}}
{\beta_\phi\gamma_\phi c\tau}\right]\,.
\label{eq:Pprompt}
\end{equation}
When $\beta_\phi\gamma_\phi c\tau\ll L_{\rm prompt}$, one finds $P_{\rm prompt}\simeq 1$, and the signal appears prompt. Conversely, for $\beta_\phi\gamma_\phi c\tau\gg L_{\rm prompt}$, the prompt rate is strongly suppressed, making displaced-vertex or invisible searches more sensitive. These probabilities naturally define three characteristic lifetime regimes:
\begin{align}
\beta_\phi\gamma_\phi c\tau \gg L_{\rm det}
&:\quad \text{escaping/invisible }\phi\,,
\nonumber\\[10pt]
L_{\rm in}\lesssim \beta_\phi\gamma_\phi c\tau \lesssim L_{\rm out}
&:\quad \text{displaced visible }\phi\,,
\nonumber\\[10pt]
\beta_\phi\gamma_\phi c\tau \ll L_{\rm prompt}
&:\quad \text{prompt visible }\phi\,.
\label{eq:three_lifetime_regimes}
\end{align}
As a result, the same point in the $(m_\phi,\sin\theta)$ plane may be probed by different experiments, depending on both the detector length scale and the boost distribution of the produced scalar. The relevant boost factor is
\begin{equation}
\beta_\phi\gamma_\phi=\frac{|\vec p_\phi^{\,\rm lab}|}{m_\phi}=\frac{\sqrt{(E_\phi^{\rm lab})^2-m_\phi^2}}{m_\phi}\,.
\label{eq:betagamma_lab}
\end{equation}
For a two-body meson decay at rest, $M\to M'\phi$, the scalar momentum is fixed by kinematics:
\begin{equation}
|\vec p_\phi^{\,*}|=\frac{\lambda^{1/2}(m_M^2,m_{M'}^2,m_\phi^2)}{2m_M}\, .
\label{eq:pstar_generic}
\end{equation}
Assuming the parent meson is approximately at rest, using Eq.~\eqref{eq:critical_mixing}, the critical mixing angle corresponding to a decay length $L$ is then,
\begin{equation}
\sin^2\theta_{c}=\frac{\lambda^{1/2}(m_M^2,m_{M'}^2,m_\phi^2)}{2m_Mm_\phi}\,
\frac{\hbar c}{L\,\Gamma_{h,\rm SM}(m_\phi)}\,.
\label{eq:critical_theta_rest_generic}
\end{equation}
This expression provides a simple estimate of the mixing angle at which the typical decay length of the scalar becomes comparable to the characteristic detector scale $L$.
\subsubsection{Case I: Detector-stable inflaton}
\label{sec:invisible}
The invisible decay constraints arise from searches for meson decays accompanied by missing energy. In each case, the limit on the mixing angle is obtained by requiring that the predicted signal rate does not exceed the experimental upper bound. The corresponding constraints are:
\begin{itemize}
\item \textbf{NA62, E787, and E949} ($K^+\to\pi^+ + {\rm inv.}$)~\cite{Atiya:1992vh,BNL-E949:2008btt}:
\begin{equation}
\st<\left[\frac{\Br^{\rm exp}_{\rm lim}(K^+\to\pi^+{\rm inv.})}{\mathcal{B}_K(\mphi)\,P_{\rm inv}}\right]^{1/2}\,,
\label{eq:K_invisible_constraint}
\end{equation}
where
\[
\Br(K^+\to\pi^+\phi)=\mathcal{B}_K(\mphi)\,\stsq\,,
\]
and $P_{\rm inv}$ denotes the probability that the scalar $\phi$ decays outside the detector volume. 
\item \textbf{KOTO} ($K_L\to\pi^0 + {\rm inv.}$)~\cite{KOTO:2014lgv,KOTO:2020prk}:
\begin{equation}
\st<\left[\frac{\Br^{\rm exp}_{\rm lim}(K_L\to\pi^0{\rm inv.})}
{\mathcal{B}_{K_L}(\mphi)\,P_{\rm inv}}
\right]^{1/2}\,,
\label{eq:KL_invisible_constraint}
\end{equation}
where
\[
\Br(K_L\to\pi^0\phi)=\mathcal{B}_{K_L}(\mphi)\,\stsq\,.
\]
\item \textbf{BaBar, Belle, and Belle\,II} ($B^+\to K^+ + {\rm inv.}$)~\cite{BaBar:2001yhh,BaBar:2010oqg,Belle:2009zue,Belle:2019oag,BelleII:2018jsg,Belle-II:2023esi}:
\begin{equation}
\st<\left[\frac{\Br^{\rm exp}_{\rm lim}(B^+\to K^+{\rm inv.})}
{\mathcal{B}_{BK}(\mphi)\,P_{\rm inv}}
\right]^{1/2}\,,
\label{eq:B_invisible_exclusive_constraint}
\end{equation}
where
\[
\Br(B^+\to K^+\phi)=\mathcal{B}_{BK}(\mphi)\,\stsq\,.
\]
\end{itemize}
\subsubsection{Case II: Detector-unstable inflaton}
If the inflaton decays into visible sector such as ($f_{\rm vis} = e^+e^-, \mu^+\mu^-,\gamma\gamma, {\rm hadrons}$), the corresponding experimental constraints must be weighted by the probability that the decay occurs within the detector region relevant to a given search. That is, for prompt searches, this probability is given by $P_{\rm reg}=P_{\rm prompt}$, whereas for displaced-vertex searches one uses $P_{\rm reg}=P_{\rm dec}$. For visible kaon decay searches, the effective signal branching ratio becomes,
\begin{equation}
{\cal B}_K(\mphi)
=
\Br(\phi\to f_{\rm vis})\,P_{\rm reg},
\end{equation}
which must satisfy,
\begin{equation}
{\cal B}_K(\mphi)
<
\Br^{\rm exp}_{\rm lim}
\!\left(K^+\to\pi^+ f_{\rm vis}\right).
\label{eq:K_visible_constraint}
\end{equation}
Similarly, constraints from $B$-meson decays are imposed through,
\begin{equation}
{\cal B}_{BK}(\mphi)
=
\Br(\phi\to f_{\rm vis})\,P_{\rm reg}
<
\Br^{\rm exp}_{\rm lim}
\!\left(B\to K f_{\rm vis}\right),
\label{eq:B_visible_constraint}
\end{equation}
where $P_{\rm reg}$ is chosen according to the search strategy. 

For far-detector experiments, the signal arises from a long-lived scalar subsequently decaying inside the detector volume. The expected number of signal events is then,
\begin{equation}
N_{\rm sig}
=
N_\phi\,
\Br(\phi\to{\rm vis.})\,
P_{\rm dec}\,
\epsilon_{\rm det},
\label{eq:far_detector_yield}
\end{equation}
where $N_\phi$ denotes the number of scalars produced within the detector acceptance and $\epsilon_{\rm det}$ is the detector efficiency. For scalars produced through $B$-meson decays,
\begin{equation}
N_\phi^{B}
\simeq
2\,{\cal L}\,\sigma_{b\bar b}\,
\Br(B\to X_s\phi),
\label{eq:Nphi_B}
\end{equation}
while production from exotic Higgs decays gives
\begin{equation}
N_\phi^{h}
=
2\,{\cal L}\,\sigma_h\,
\Br(h\to\phi\phi).
\label{eq:Nphi_h}
\end{equation}
To derive projected sensitivities, one typically requires a minimum number of signal events. For MATHUSLA, the usual background-free sensitivity convention is $N_{\rm sig}\geq 4$~\cite{MATHUSLA:2018bqv}. For SHiP and FASER projections, one often uses $N_{\rm sig}\geq2.3$ for a 90\% CL zero-background exclusion~\cite{SHiP:2020vbd, FASER:2018eoc}.
\subsection{Constraints from neutral meson oscillations }
Beyond the direct on-shell production channels discussed above, the scalar mixing
also generates flavour-changing neutral current (FCNC) effects at the loop level.
The same Higgs-penguin diagram responsible for $q_i\!\to\! q_j\phi$ emission, when
the $\phi$ propagator is made virtual rather than put on-shell, contributes to
$K^0$--$\bar{K}^0$ and $B^0$--$\bar{B}^0$ oscillations through the $|\Delta F|=2$ effective operator. This provides an indirect constraint on the mixing angle that does not require $\phi$ to be produced or detected, and that is therefore independent of the decay length. 

Integrating out the virtual $\phi$ at tree level generates the scalar-scalar
$|\Delta F|=2$ operator,
\begin{equation}
Q_4 = (\bar{q}_L q^{\prime}_{R})(\bar{q}^{\prime}_{R} q_{ L})\,,
\label{eq:Q4}
\end{equation}
with Wilson coefficient, 
\begin{equation}
C_4^\phi=\frac{|C_{qq^{\prime}}|^2}{2\,m_\phi^2}\,,
\label{eq:C4}
\end{equation}
where $C_{qq^{\prime}}$ is given by the Higgs-penguin matching similar to the one given in Eq.\eqref{eq:Cbs}.
Note that the factor $V_{tq}^*V_{tq^{\prime}}$ is much larger for $B^0$--$\bar{B}^0$ than $K^0$--$\bar{K}^0$ oscillation and, hence, the constraints from $B^0$--$\bar{B}^0$ oscillation dominates over the Kaon oscillation by a factor $(m_b|V_{ts}V_{tb}|)/(m_s|V_{ts}V_{td}|) \approx 5\times10^{3}$. Thus, we will consider only the $|\Delta F=2|$, $B^0$--$\bar{B}^0$ oscillation process, with $q = s$ and $q'=b$ in Eq.\eqref{eq:Q4}. The operator $Q_4$ is chirality-flipping on both quark lines, which leads to a hadronic matrix element that is chirally enhanced, making this one of the most tightly constrained operators in the $\Delta F=2$
sector, as emphasized in the UTfit model-independent analysis~\cite{Bona:2007vi}.


Following the standard convention used in lattice QCD calculations and adopting the FLAG averages~\cite{FLAGReview:2021npn}, the hadronic matrix element of the operator $Q_4$ can be parametrized as,
\begin{equation}
\langle \bar{B}^0 | Q_4 | B^0 \rangle
=
\frac{5}{3}\,B_4
\left(\frac{m_B}{m_b+m_s}\right)^2
\frac{f_B^2\,m_B}{4}\, ,
\label{eq:me4}
\end{equation}
where, $m_B = 5.279\,\mathrm{GeV}$, $m_b = 4.18\,\mathrm{GeV}$, and $m_s = 0.093\,\mathrm{GeV}$ ($\overline{\rm MS}$ scheme at $2\,\mathrm{GeV}$).
Using the bag parameter $B_4 = 0.78$ and the decay constant $f_B = 0.190\,\mathrm{GeV}$~\cite{Dowdall:2019bea}, we obtain,
\begin{equation}
\langle \bar{B}^0 | Q_4 | B^0 \rangle
=
9.5 \times 10^{-2}\,\mathrm{GeV}^3\,.
\end{equation}

Requiring, $\Big|\frac{C_4^\phi\,\langle\bar{B}^0|Q_4|B^0\rangle}{2\,m_B}\Big| \leq\Delta m_B^\mathrm{exp}/2$, where $\Delta m_B^\mathrm{exp} = 3.334\times10^{-13}\;\mathrm{GeV}$~\cite{ParticleDataGroup:2022pth} is the mass splitting, we obtain,
\begin{equation}
C_4^\phi\;\leq\;C_4^\mathrm{max}\;\equiv\;\frac{\Delta m_B^\mathrm{exp}\,m_B}{\langle \bar{B}^0 | Q_4 | B^0 \rangle}=1.86\times10^{-11}\;\mathrm{GeV}^{-2}\,.
\label{eq:C4max}
\end{equation}
Substituting Eq.~\eqref{eq:Cbs} into Eq.~\eqref{eq:C4max} and using the expression for $C_{bs}$ in Eq.\eqref{eq:Cbs} the solution for $\sin\theta$ becomes,
\begin{equation}
\sin\theta\;<\;\left(\frac{2\,C_4^\mathrm{max}}{|C_{bs}/\sin\theta|^2}\right)^{1/2}\!\! m_\phi\;=\;\alpha_{\Delta F}\,m_\phi\,,
\label{eq:DmB_bound}
\end{equation}
with
\begin{equation}
\alpha_{\Delta F}=\frac{\sqrt{2\,C_4^\mathrm{max}}}{|C_{bs}/\sin\theta|}=\frac{16\pi^2v_h^3\sqrt{2\,C_4^\mathrm{max}}}   {3\,m_b\,m_t^2\,|V_{ts}^*V_{tb}|} \approx 0.96\;\mathrm{GeV}^{-1}\,.
  \label{eq:alpha_DmB}
\end{equation}
The bound is therefore linear in $m_\phi$, arising because $C_4^\phi
\propto m_\phi^{-2}$ while $\sin^2\theta$ enters linearly in $C_4^\phi$. It becomes vacuous ($\sin\theta > 1$) above
$m_\phi^\mathrm{max} = 1/\alpha_{\Delta F} \approx 1.04\;\mathrm{GeV}$, and at
$m_\phi = 1\;\mathrm{MeV}$ gives $\sin\theta < 9.6\times10^{-4}$.
\subsection{Summary of laboratory bounds}
\label{sec:expts}
\begin{table}[htb!]
\centering
\begin{tabular}{c|c|c|c} \hline
Experiment & Production channel & Detector geometry & Constraint (Br) \\ \hline
E787/E949 & $K^+ \to \pi^+\phi $ & $L_{\rm det}=1$m \cite{Atiya:1992vh} & $<7.3\times10^{-11}$ \cite{BNL-E949:2008btt} \\
NA62 & $K^+ \to \pi^+\phi $ & $L_{\rm det}=120$m \cite{NA62:2017rwk} & $<1.06\times10^{-10}$ \cite{NA62:2021dod} \\
KOTO & $K_L \to \pi^0\phi $ & $L_{\rm det}=3$m \cite{KOTO:2014lgv} & $<3.0\times10^{-9}$ \cite{KOTO:2020prk} \\
BaBar & $B^+ \to K^+\phi $ & $L_{\rm det}=1.5$m \cite{BaBar:2001yhh}& $<1.9\times10^{-5}$ \cite{BaBar:2010oqg} \\
Belle II & $B^+ \to K^+\phi $ & $L_{\rm det}=1.0$m \cite{BelleII:2018jsg}& $<4.0\times10^{-6}$ \cite{Belle-II:2023esi}\\ \hline
\end{tabular}
\caption{Experimental constraint and detector geometry used to constraint a {\it detector-stable} Higgs-mixed light inflaton. The signal rate is $\mathrm{Br}_{\rm prod}\times\sin^2 \theta\times P_{\rm inv}$, where $P_{\rm inv}=\exp(-L_{\rm det}/\beta\gamma c\tau_\phi)$ is the probability for $\phi$ to escape the detector and with $c\tau_\phi = \hbar c/(\sin^2 \theta\,\Gamma_h^{\rm SM}(m_\phi))$.}
\label{tab:invisible}
\end{table}
\begin{table}[htb!]
\centering
\begin{tabular}{c|c|c|c} \hline
Experiment & Production channel & Detector geometry & Constraint (Br) \\ \hline
NA62      & $K^+\to\pi^+\phi,\;\phi\to\mu^+\mu^-$    & $L\in[10\,\text{m},120\,\text{m}]$\cite{NA62:2017rwk} & $<9.15\times10^{-9}$ \cite{NA62:2021dod} \\
BaBar/Belle & $B\to K\phi,\;\phi\to\mu^+\mu^-$       & $L\in[1\,\text{cm},1.5\,\text{m}]$ \cite{BaBar:2001yhh}& $<2.5\times10^{-7}$ \cite{Belle:2019oag} \\
LHCb  & $B\to K\phi,\;\phi\to\mu^+\mu^-$    & $L\in[2\,\text{mm},10\,\text{m}]$ \cite{LHCb:2008vvz}& $<3\times10^{-9}$ \cite{LHCb:2016awg} \\
MATHUSLA  & $B\to X_s\phi$ (incl.), $\phi\to\text{vis.}$ & $L\in[200\,\text{m},220\,\text{m}]$ \cite{Curtin:2018mvb} & $N_{\rm sig}\geq4$ \cite{Curtin:2018mvb} \\
FASER2    & $B\to K\phi,\;\phi\to\text{vis.}$         & $L\in[480\,\text{m},485\,\text{m}]$ \cite{FASER:2018eoc} & $N_{\rm sig}\geq2.3$ \cite{FASER:2018eoc} \\
SHiP  & $K\to\pi\phi$ or $B\to X_s\phi$, $\phi\to\text{vis.}$ & $L\in[45\,\text{m},120\,\text{m}]$ \cite{Miano:2812714, SHiP:2021nfo} & $N_{\rm sig}\geq2.3$ \cite{SHiP:2021nfo} \\ \hline
\end{tabular}
\caption{Experimental constraint and detector geometry used to constraint a {\it detector-unstable} Higgs-mixed light inflaton. The signal rate is $\mathrm{Br}_{\rm prod}\times\sin^2\!\theta \times\mathrm{Br}(\phi\to f_{\rm vis})\times P_{\rm reg}$, where $P_{\rm reg}=P_{\rm prompt}=1-e^{-L_p/\beta\gamma c\tau_\phi}$ for prompt searches and $P_{\rm reg}=P_{\rm dec}=e^{-L_{\rm in}/\beta\gamma c\tau_\phi}-e^{-L_{\rm out}/\beta\gamma c\tau_\phi}$ for displaced-vertex searches, with $c\tau_\phi=\hbar c/(\sin^2\!\theta\,\Gamma_h^{\rm SM}(m_\phi))$. The projected sensitivities are from MATHUSLA, FASER2, and SHiP.}
\label{tab:visible}
\end{table}
\begin{figure}[htb!]
    \centering    \includegraphics[scale=0.7]{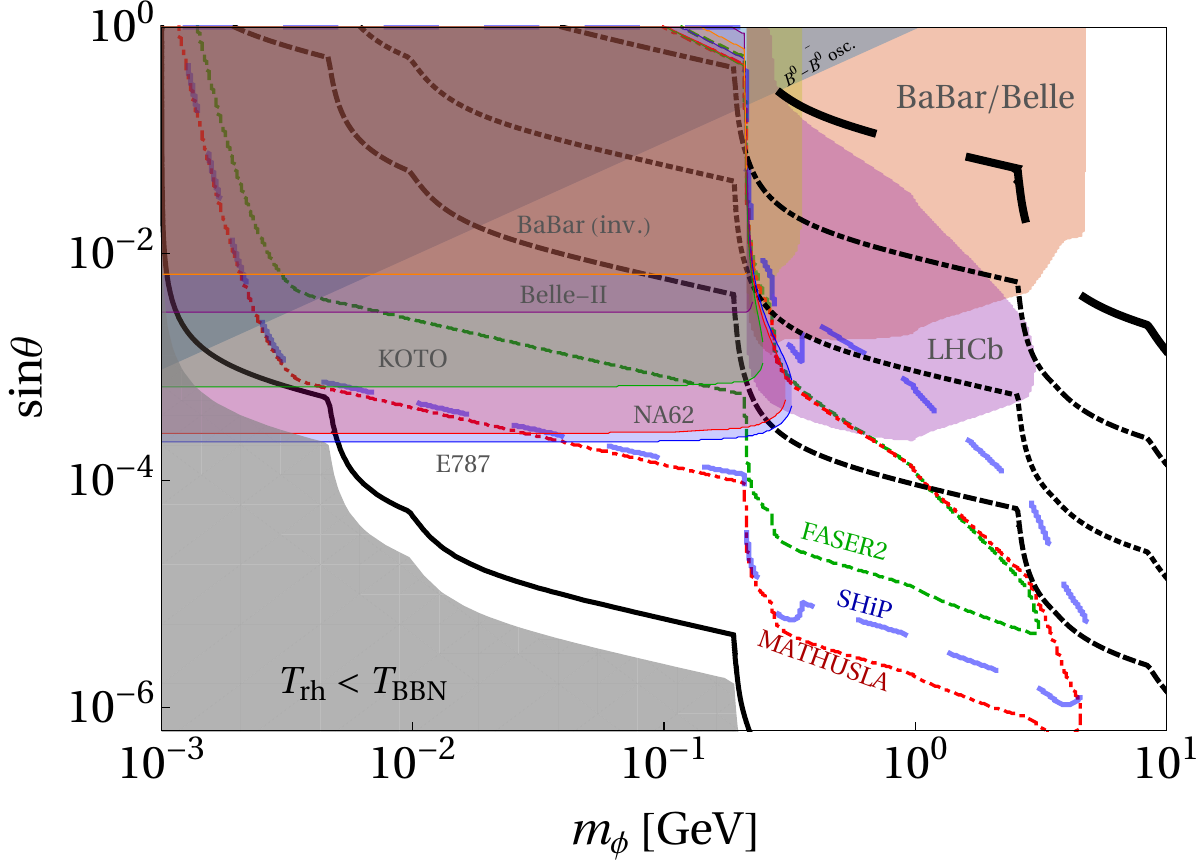}
    \caption{Exclusion plot for a light inflaton. The solid, dashed, dotted dot-dashed and thick-dashed black contours correspond to $\Trh=\{10^{-2},\,10,\,100,\,10^3,\,10^4\}$ GeV, respectively. The bottom left gray shaded region is disallowed due to lower bound on $\Trh$ from BBN. Different coloured shaded regions are bounds obtained from detector-stable and detector-unstable inflaton search experiments, as tabulated in Tab.~\ref{tab:invisible} and Tab.~\ref{tab:visible}, respectively.}    
    \label{fig:treh}
\end{figure}
The experiments used to constrain the mixing angle $\st$ as a function of the inflaton mass $\mphi$ are listed in Tab.~\ref{tab:invisible} and Tab.~\ref{tab:visible}, corresponding to searches for detector-stable and detector-unstable inflatons, respectively. Searches such as NA62~($K^+\!\to\pi^+\mu^+\mu^-$) and BaBar/Belle~($B\!\to K\mu^+\mu^-$) require $\phi$ to decay within centimeters of the production vertex. At large $\sin\theta$ the lifetime is short and $P_\mathrm{prompt}\to 1$, so the signal grows monotonically with the mixing angle. These searches therefore produce a simple lower bound on $\sin\theta$. Whereas, the LHCb search for $B\!\to  \phi(\to\mu^+\mu^-)$ with displaced vertices uses the fiducial volume $L_\mathrm{in}=2\;\mathrm{mm}$, from the VErtex LOcator (VELO), to $L_\mathrm{out}=10\;\mathrm{m}$. This probability peaks when $\beta_{\phi}\gamma_{\phi}\,c\tau_\phi \sim L_\mathrm{in}$–$L_\mathrm{out}$ and vanishes in both the short and long lifetime limits. Consequently, the LHCb exclusion forms a closed island in the $(m_\phi,\sin\theta)$ plane, with the upper boundary arising because of prompt inflaton decay at sufficiently large  $\sin\theta$, within 2\,mm displacement of VELO, thus falling outside the LHCb displaced-vertex acceptance. MATHUSLA, FRASER2 and SHiP, on the other hand, have fiducial regions beginning hundreds of metres from the interaction point ($L_\mathrm{in}=200\;\mathrm{m}$ for MATHUSLA, $L_\mathrm{in}=480\;\mathrm{m}$ for FRASER2, and $L_\mathrm{in}=45\;\mathrm{m}$ for SHiP) and hence, are sensitive only to long-lived scalars with small mixing angle requiring $\beta_{\phi}\gamma_{\phi}\,c\tau_\phi \gtrsim L_\mathrm{in}$. Their sensitivity contour therefore extends to very small $\sin\theta$, probing parameter space inaccessible to any near-detector experiment. 

At NA62~($K^+\!\to\pi^+ \ell^+\ell^-$) and BaBar/Belle~($B\!\to K\ell^+\ell^-$), search for the Higgs mixed scalar decaying to light leptons is strongly disfavoured due to their small Yukawa. As a consequence, $\mathrm{Br}(\phi\!\to e^+e^-)$ is suppressed by a factor of $\sim\!40,000$ relative to $\mathrm{Br}(\phi\!\to\mu^+\mu^-)$, making the electron channel negligible for prompt and displaced searches whose signal yield is directly proportional to $\mathrm{Br}(\phi\to f_\mathrm{vis})$. The same helicity suppression that renders $\phi\!\to e^+e^-$ useless for near-detector searches makes it ideal for MATHUSLA and SHiP. The extremely small electron-channel width means that even at mixing angles as large as $\sin\theta\sim 10^{-3}$, the proper decay length satisfies $\beta\gamma\,c\tau_\phi\gg 1\;\mathrm{m}$. This complementarity means that MATHUSLA and SHiP are not simply extending the reach of near-detector searches in sensitivity, rather, they are the only experiments probing the $2m_e < m_\phi < 2m_\mu$ window, where the inflaton's extreme longevity in the electron-only regime could be searched for. For $m_\phi \lesssim$ MeV, where the kaon and $B$-meson direct-decay searches lose sensitivity, because of the kinematic inaccessibility or detector-stability of the inflaton, the $\Delta m_B$ constraint from $B^0$--$\bar{B}^0$ oscillations become important. The significance of this bound lies in the fact that it is independent of the inflaton lifetime, rather, it constrains $\sin\theta$ regardless of whether $\phi$ decays promptly, displaced, or escapes entirely.

The sensitivity of current and future searches for a detector-(un)stable inflaton, and their implications for the reheating temperature, is illustrated in Fig.~\ref{fig:treh}. The nature of the curves that result in different $\Trh$, are easy to understand from the fact that $\Trh\propto\theta^2\,m_\phi$, following Eq.~\eqref{eq:treh-gamma}. Existing constraints from $B^0-\bar{B}^0$ oscillations, BaBar/Belle, KOTO, NA62, E787 and LHCb already exclude a significant portion of the parameter space. Future experiments such as FASER2, SHiP, and MATHUSLA will extend the sensitivity to lower reheating temperatures $\Trh\lesssim 1$ GeV. In addition, BBN imposes a stronger constraint, excluding the region $m_\phi\lesssim 0.1$ GeV, where $T_{\rm BBN}\simeq 4$ MeV, depending on the mixing angle. For any given value of $\Trh$, the corresponding values of $\xi_\phi$ and $\Delta_L$ required to satisfy the ACT data can be directly inferred from Fig.~\ref{fig:Nk-Trh}. 
\section{Implications for dark matter}
\label{sec:dm}
Up to this point, we have assumed that reheating transfers the full inflaton energy density to the visible sector. We now revoke this assumption and consider the possibility that part of the inflaton energy density is channeled into physics beyond the SM. We assume that the inflaton can also decay invisibly into dark matter (DM) final states with a certain branching fraction $\br$, which can be constrained by the requirement of reproducing the observed DM abundance at the end of reheating. Given a particular DM spin and its interaction with the inflaton, one can obtain an exact expression for $\br$. Here, we simply consider this to be a free parameter to make the analysis model-independent. With this, the DM production rate can be obtained by solving BEQ of the form,
\begin{align}
& \frac{d\Ndm}{da}=\frac{a^2\,\gamma_{\phi\to\text{DM}}}{\mathcal{H}}\,,    
\end{align}
where $\Ndm=\ndm\,a^3$ denotes the comoving DM number, and 
\begin{align}
& \gamma_{\phi\to\text{DM}} = 2\,\br\,\Gamma_\phi\,\frac{\rho_\phi}{m_\phi}\,,    
\end{align}
where the factor of 2 takes into account DM pair-production. If DM is produced during reheating $(\aend\ll a\ll\arh)$, the above equation admits an analytical solution, leading to the DM yield,
\begin{align}\label{eq:Y0}
& Y(\arh)\equiv Y_0=\frac{\ndm(\arh)}{s(\arh)}\simeq\frac{9\sqrt{\pi}\,\br}{2}\,\left(\frac{5}{2}\right)^{1/4}\,\frac{\gs(\Trh)^{3/4}}{\gss(\Trh)}\,\sqrt{\frac{M_P\,\Gamma_\phi}{m_\phi^2}}\,,
\end{align}
which corresponds to the present DM abundance $Y_0$, considering no further entropy injection from the end of reheating until today. Naively, since $\Gamma_\phi\sim\theta^2\,m_\phi$, hence the final yield becomes proportional to $\sqrt{\theta^2\,m_\phi}$. Consequently, the final DM abundance is determined by the following set of independent parameters: $$\{\mdm,\,m_\phi,\,\theta,\,\br\}\,.$$ Reproducing the observed DM relic density requires $Y_0\, \mdm = \Omega h^2 \, \frac{1}{s_0}\,\frac{\rho_c}{h^2} \simeq 4.3 \times 10^{-10}\,\text{ GeV}$. Here we use the critical energy density $\rho_c \simeq 1.05 \times 10^{-5}\, h^2$~GeV/cm$^3$, the present entropy density $s_0\simeq 2.69 \times 10^3$~cm$^{-3}$~\cite{ParticleDataGroup:2022pth}, and the DM relic abundance $\Omega h^2 \simeq 0.12$, with $h\simeq H_0/100\,\left({\rm km/s/Mpc}\right)$ the reduced Hubble parameter, where $H_0\simeq 67.4 \pm 0.5 \text{km/s/Mpc}$ is the present Hubble rate~\cite{Planck:2018vyg}.  
\begin{figure}[htb!]
\centering
\includegraphics[scale=0.7]{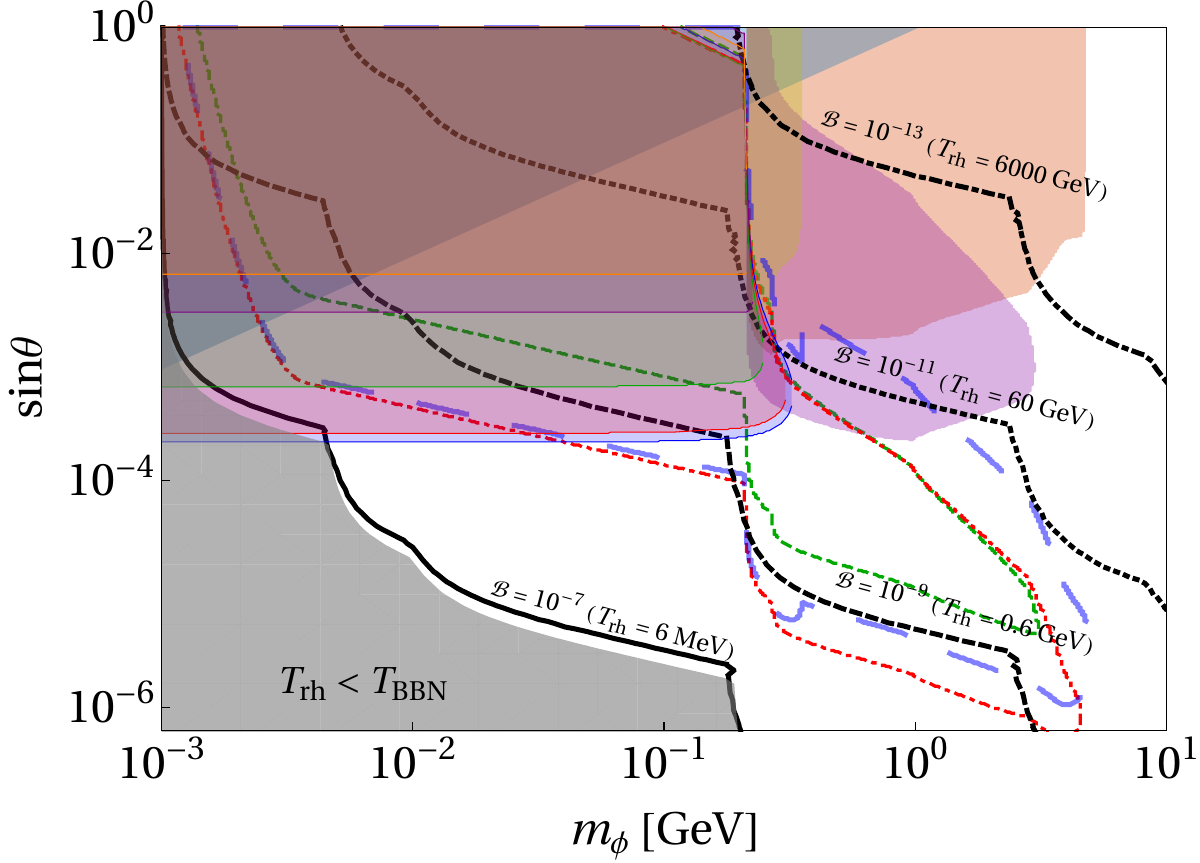}
\caption{Contours of right relic abundance, for $\mdm=\mphi/3$, considering different choices of the branching fraction as shown via different black curves. The exclusion limits are same as in Fig.~\ref{fig:treh}.}
\label{fig:rel}
\end{figure}

The contours satisfying observed relic abundance, for different choices of $\br$, are shown in Fig.~\ref{fig:rel}, considering DM production entirely from inflaton decay. We have fixed the DM mass to be $\mdm=m_\phi/3$, as a benchmark value. Note that, for a given $m_\phi$, a smaller branching requires larger $\theta$, which follows from Eq.~\eqref{eq:Y0}. This can be understood from the fact that a smaller $\br$ reduces the DM production per inflaton decay, which must be compensated by a larger total inflaton decay width, $\Gamma_\phi\propto \theta^2$, thereby increasing the overall number of inflaton decays during reheating. As a result, corresponding $\Trh$ also increases with smaller $\br$, as one can see from the figure. Different laboratory searches (same as in Fig.~~\ref{fig:treh}) typically put tight constraints on larger $\br$, and for $\sin\theta\gtrsim 10^{-3}$, while smaller $\br$'s are rather unconstrained. One could also produce DM via $\phi$-mediated scattering, however such production cross-sections will be sub-dominant due to $\theta^2\br^2$ suppression. For each value of $\Trh$, the corresponding values of $\xi_\phi$ and $\Delta_L$ that satisfy the ACT constraints can be obtained from Figs.~\ref{fig:nsvr} and \ref{fig:Nk-Trh}. Together, it provides a self-consistent picture of both SM and DM production during reheating after inflation.
\section{Conclusions}
\label{sec:concl}
Low energy intensity-frontier experiments provide a powerful avenue for searching for feebly interacting light particles beyond the Standard Model. If such light new physics is produced in the early Universe before big bang nucleosynthesis (BBN), these experiments can shed light on the otherwise inaccessible pre-BBN era. An especially compelling scenario is when the light state is the inflaton itself. In this case, terrestrial experiments could directly search for the inflaton, providing a unique opportunity to test the physics that drove the exponential expansion of the early Universe. This would provide an indirect probe of inflation, complementing the conventional approach based on the yet-to-be unambiguously detected stochastic primordial gravitational wave background. In this work, we revisit the scale-invariant quartic inflationary framework proposed in the Light Inflaton Hunter's Guide~\cite{Bezrukov:2009yw} and confront it with the latest CMB constraints from ACT. Focusing on inflaton masses $\lesssim 10 \,\mathrm{GeV}$, we perform a comprehensive study of the discovery potential of current and future laboratory experiments, including NA62, KOTO, BaBar, Belle, LHCb, SHiP, FASER2, and MATHUSLA. We identify the regions of parameter space that simultaneously yield successful reheating and satisfy inflationary constraints. Finally, we demonstrate that the same framework also accommodates the production of particle dark matter with arbitrary spin during reheating, accounting for the observed relic abundance.  
\section*{Acknowledgments}
The authors would like to thank IIT, Hyderabad for their hospitality during WHEPP-2025, where this work was initiated during working group discussion sessions. The authors would also like to thank Dmitry Gorbunov for providing valuable feedback on the draft. 
\bibliography{Bibliography}

@article{Atiya:1992vh,
  author        = {Atiya, M. S. and others},
  collaboration = {E787},
  title         = {{The E787 detector for the study of rare kaon decays}},
  journal       = {Nucl. Instrum. Meth. A},
  volume        = {321},
  year          = {1992},
  pages         = {129},
  doi           = {10.1016/0168-9002(92)90379-J},
  note          = {E787/E949 detector paper. Stopped kaon technique,
                   $p_K\approx0.71$\,GeV/$c$; range-stack outer radius $\sim$1\,m}
}

@article{NA62:2017rwk,
  author        = {Cortina Gil, E. and others},
  collaboration = {NA62},
  title         = {{The beam and detector of the NA62 experiment at CERN}},
  journal       = {JINST},
  volume        = {12},
  year          = {2017},
  pages         = {P05025},
  doi           = {10.1088/1748-0221/12/05/P05025},
  eprint        = {1703.08501},
  archivePrefix = {arXiv},
  primaryClass  = {physics.ins-det},
  note          = {NA62 detector TDR. 75\,GeV/$c$ $K^+$ beam;
                   120\,m fiducial vacuum decay region}
}

@article{Dowdall:2019bea,
    author = "Dowdall, R. J. and Davies, C. T. H. and Horgan, R. R. and Lepage, G. P. and Monahan, C. J. and Shigemitsu, J. and Wingate, M.",
    title = "{Neutral B-meson mixing from full lattice QCD at the physical point}",
    eprint = "1907.01025",
    archivePrefix = "arXiv",
    primaryClass = "hep-lat",
    reportNumber = "INT-PUB-19-031, JLAB-THY-19-3068",
    doi = "10.1103/PhysRevD.100.094508",
    journal = "Phys. Rev. D",
    volume = "100",
    number = "9",
    pages = "094508",
    year = "2019"
}

@article{Bona:2007vi,
    author = "Bona, M. and others",
    collaboration = "UTfit",
    title = "{Model-independent constraints on $\Delta F=2$ operators and the scale of new physics}",
    eprint = "0707.0636",
    archivePrefix = "arXiv",
    primaryClass = "hep-ph",
    doi = "10.1088/1126-6708/2008/03/049",
    journal = "JHEP",
    volume = "03",
    pages = "049",
    year = "2008"
}

@article{FLAGReview:2021npn,
    author = "Aoki, Y. and others",
    collaboration = "Flavour Lattice Averaging Group (FLAG)",
    title = "{FLAG Review 2021}",
    eprint = "2111.09849",
    archivePrefix = "arXiv",
    primaryClass = "hep-lat",
    reportNumber = "CERN-TH-2021-191, JLAB-THY-21-3528, FERMILAB-PUB-21-620-SCD-T",
    doi = "10.1140/epjc/s10052-022-10536-1",
    journal = "Eur. Phys. J. C",
    volume = "82",
    number = "10",
    pages = "869",
    year = "2022"
}

@article{KOTO:2014lgv,
  author        = {Ahn, J. K. and others},
  collaboration = {KOTO},
  title         = {{Search for the $K_L\to\pi^0\nu\bar{\nu}$ and
                   $K_L\to\pi^0 X^0$ decays at the J-PARC KOTO experiment}},
  journal       = {Phys. Rev. Lett.},
  volume        = {122},
  year          = {2019},
  pages         = {021802},
  doi           = {10.1103/PhysRevLett.122.021802},
  eprint        = {1810.09710},
  archivePrefix = {arXiv},
  primaryClass  = {hep-ex},
  note          = {KOTO detector description: 3\,m $K_L$ decay volume,
                   CsI calorimeter at downstream end, $p_{K_L}\approx2$\,GeV/$c$}
}

@article{BaBar:2001yhh,
  author        = {Aubert, B. and others},
  collaboration = {BaBar},
  title         = {{The BaBar detector}},
  journal       = {Nucl. Instrum. Meth. A},
  volume        = {479},
  year          = {2002},
  pages         = {1},
  doi           = {10.1016/S0168-9002(01)02012-5},
  eprint        = {hep-ex/0105044},
  archivePrefix = {arXiv},
  primaryClass  = {hep-ex},
  note          = {BaBar detector paper. EMC outer radius $\approx$1.36\,m;
                   IFR extends to $\approx$1.75\,m.
                   Used for BaBar visible and invisible searches}
}

@article{BelleII:2018jsg,
  author        = {Kou, E. and others},
  collaboration = {Belle II},
  title         = {{The Belle II physics book}},
  journal       = {PTEP},
  volume        = {2019},
  year          = {2019},
  pages         = {123C01},
  doi           = {10.1093/ptep/ptz106},
  eprint        = {1808.10567},
  archivePrefix = {arXiv},
  primaryClass  = {hep-ex},
  note          = {Belle II TDR. KLM outer radius $\approx$1.0\,m;
                   $B$ produced near rest at $\Upsilon(4S)$}
}

@article{LHCb:2008vvz,
  author        = {Alves, A. A. and others},
  collaboration = {LHCb},
  title         = {{The LHCb detector at the LHC}},
  journal       = {JINST},
  volume        = {3},
  year          = {2008},
  pages         = {S08005},
  doi           = {10.1088/1748-0221/3/08/S08005},
  note          = {LHCb detector paper. VELO from $-0.3$\,m to $+0.7$\,m from IP;
                   downstream spectrometer to $\sim$9\,m.
                   Displaced-vertex fiducial $L\in[2\,\mathrm{mm},10\,\mathrm{m}]$}
}

@article{Curtin:2018mvb,
  author        = {Curtin, David and others},
  title         = {{Long-Lived Particles at the Energy Frontier:
                   The MATHUSLA Physics Case}},
  journal       = {Rept. Prog. Phys.},
  volume        = {82},
  year          = {2019},
  pages         = {116201},
  doi           = {10.1088/1361-6633/ab28d6},
  eprint        = {1806.07396},
  archivePrefix = {arXiv},
  primaryClass  = {hep-ph},
  note          = {MATHUSLA design and physics case.
                   Decay volume $L\in[200,220]$\,m from IP;
                   $N_{\rm sig}\geq4$, $N_B^{\rm HL}\varepsilon=5\times10^{14}$}
}

@article{FASER:2018eoc,
  author        = {Ariga, A. and others},
  collaboration = {FASER},
  title         = {{Detecting and studying high-energy collider neutrinos
                   with FASER at the LHC}},
  journal       = {Phys. Rev. D},
  volume        = {99},
  year          = {2019},
  pages         = {095011},
  doi           = {10.1103/PhysRevD.99.095011},
  eprint        = {1811.10243},
  archivePrefix = {arXiv},
  primaryClass  = {hep-ph},
  note          = {FASER/FASER2 design. Location 480\,m from ATLAS IP;
                   FASER2 decay volume $\sim$5\,m length.
                   $N_{\rm sig}\geq2.3$, $N_B^{\rm fwd}\varepsilon=5\times10^{13}$}
}

@article{SHiP:2021nfo,
  author        = {Ahdida, C. and others},
  collaboration = {SHiP},
  title         = {{The SHiP experiment at the proposed CERN SPS Beam Dump Facility}},
  journal       = {JHEP},
  volume        = {04},
  year          = {2021},
  pages         = {199},
  doi           = {10.1007/JHEP04(2021)199},
  eprint        = {2002.12150},
  archivePrefix = {arXiv},
  primaryClass  = {physics.ins-det},
  note          = {SHiP TDR. Decay volume $L\in[30,100]$\,m from target;
                   $2\times10^{20}$ POT at 400\,GeV SPS,
                   $N_K=3\times10^{17}$, $N_B=3\times10^{13}$.
                   $N_{\rm sig}\geq2.3$}
}

@article{Belle:2009zue,
  author        = {Wei, J.-T. and others},
  collaboration = {Belle},
  title         = {{Measurement of the differential branching fraction and
                   forward-backward asymmetry for $B\to K^{(*)}\ell^+\ell^-$}},
  journal       = {Phys. Rev. Lett.},
  volume        = {103},
  year          = {2009},
  pages         = {171801},
  doi           = {10.1103/PhysRevLett.103.171801},
  eprint        = {0904.0770},
  archivePrefix = {arXiv},
  primaryClass  = {hep-ex},
  note          = {$\mathcal{B}(B\to K\ell^+\ell^-)=(4.8^{+0.5}_{-0.4}\pm0.3)\times10^{-7}$;
                   used as the constraint for visible $\phi\to\mu^+\mu^-$ row.
                   NOTE: previously incorrectly cited as arXiv:2009.00512,
                   which is an $R_K$ ratio measurement, not a branching fraction limit}
}

@article{LHCb:2016awg,
  author        = {Aaij, R. and others},
  collaboration = {LHCb},
  title         = {{Search for long-lived scalar particles in
                   $B^+\to K^+\chi(\mu^+\mu^-)$ decays}},
  journal       = {Phys. Rev. D},
  volume        = {95},
  year          = {2017},
  pages         = {071101},
  doi           = {10.1103/PhysRevD.95.071101},
  eprint        = {1612.07818},
  archivePrefix = {arXiv},
  primaryClass  = {hep-ex},
  note          = {$\mathcal{B}(B\to K\mu^+\mu^-)_{\rm DV}<3\times10^{-9}$;
                   displaced dimuon in VELO $L\in[2\,\mathrm{mm},10\,\mathrm{m}]$}
}

@article{BNL-E949:2008btt,
  author        = {Anisimovsky, V. V. and others},
  collaboration = {BNL-E949},
  title         = {{Study of $K^+\to\pi^+\nu\bar{\nu}$ in the momentum region
                   $140<P_\pi<199$\,MeV/$c$}},
  journal       = {Phys. Rev. Lett.},
  volume        = {101},
  year          = {2008},
  pages         = {191802},
  doi           = {10.1103/PhysRevLett.101.191802},
  eprint        = {0808.2459},
  archivePrefix = {arXiv},
  primaryClass  = {hep-ex},
  note          = {Combined E787+E949 result.
                   $\mathcal{B}(K^+\to\pi^++\mathrm{inv.})<7.3\times10^{-11}$}
}

@article{NA62:2021dod,
  author        = {Cortina Gil, E. and others},
  collaboration = {NA62},
  title         = {{Measurement of the very rare $K^+\to\pi^+\nu\bar{\nu}$ decay}},
     doi = "10.1007/JHEP06(2021)093",
    journal = "JHEP",
    volume = "06",
    pages = "093",
    year = "2021",
  eprint        = {2103.15389},
  archivePrefix = {arXiv},
  primaryClass  = {hep-ex},
  note          = {$\mathcal{B}(K^+\to\pi^++\mathrm{inv.})<1.06\times10^{-10}$ at 95\% CL;
                   gap at 110--155\,MeV.
                   NOTE: arXiv number is 2103.15389 (not 2104.12336,
                   which is a different NA62 paper)}
}

@article{Belle:2019oag,
    author = "Abdesselam, A. and others",
    collaboration = "Belle",
    title = "{Test of Lepton-Flavor Universality in ${B\to K^\ast\ell^+\ell^-}$ Decays at Belle}",
    eprint = "1904.02440",
    archivePrefix = "arXiv",
    primaryClass = "hep-ex",
    reportNumber = "BELLE-CONF-1901, Belle Preprint 2020-14, KEK Preprint 2020-16",
    doi = "10.1103/PhysRevLett.126.161801",
    journal = "Phys. Rev. Lett.",
    volume = "126",
    number = "16",
    pages = "161801",
    year = "2021"
}

@article{BaBar:2010oqg,
    author = "del Amo Sanchez, P. and others",
    collaboration = "BaBar",
    title = "{Search for the Rare Decay $B \to K \nu \bar{\nu}$}",
    eprint = "1009.1529",
    archivePrefix = "arXiv",
    primaryClass = "hep-ex",
    reportNumber = "SLAC-PUB-14237, BABAR-PUB-10-011",
    doi = "10.1103/PhysRevD.82.112002",
    journal = "Phys. Rev. D",
    volume = "82",
    pages = "112002",
    year = "2010"
}

@article{KOTO:2020prk,
  author        = {Iwai, E. and others},
  collaboration = {KOTO},
  title         = {{Search for the $K_L\to\pi^0\nu\bar{\nu}$ and $K_L\to\pi^0 X^0$
                   decays at the J-PARC KOTO experiment}},
  journal       = {Phys. Rev. Lett.},
  volume        = {126},
  year          = {2021},
  pages         = {121801},
  doi           = {10.1103/PhysRevLett.126.121801},
  eprint        = {2012.07301},
  archivePrefix = {arXiv},
  primaryClass  = {hep-ex},
  note          = {$\mathcal{B}(K_L\to\pi^0+\mathrm{inv.})<3.0\times10^{-9}$
                   at 90\% CL}
}

@article{Belle-II:2023esi,
  author        = {Adachi, I. and others},
  collaboration = {Belle II},
  title         = {{Evidence for $B^+\to K^+\nu\bar{\nu}$ decays}},
  journal       = {Phys. Rev. Lett.},
  volume        = {131},
  year          = {2023},
  pages         = {051803},
  doi           = {10.1103/PhysRevLett.131.051803},
  eprint        = {2303.12274},
  archivePrefix = {arXiv},
  primaryClass  = {hep-ex},
  note          = {$\mathcal{B}(B^+\to K^++\mathrm{inv.})<4.0\times10^{-6}$;
                   first hadronic-tag result from Belle II}
}

@article{MATHUSLA:2018bqv,
    author = "Alpigiani, Cristiano and others",
    collaboration = "MATHUSLA",
    title = "{A Letter of Intent for MATHUSLA: A Dedicated Displaced Vertex Detector above ATLAS or CMS.}",
    eprint = "1811.00927",
    archivePrefix = "arXiv",
    primaryClass = "physics.ins-det",
    reportNumber = "CERN-LHCC-2018-025, LHCC-I-031",
    month = "7",
    year = "2018"
}

@article{SHiP:2020vbd,
    author = "Ahdida, C. and others",
    collaboration = "SHiP",
    title = "{Sensitivity of the SHiP experiment to dark photons decaying to a pair of charged particles}",
    eprint = "2011.05115",
    archivePrefix = "arXiv",
    primaryClass = "hep-ex",
    doi = "10.1140/epjc/s10052-021-09224-3",
    journal = "Eur. Phys. J. C",
    volume = "81",
    number = "5",
    pages = "451",
    year = "2021"
}

@article{Miano:2812714,
      author        = "Miano, Andrea and Fiorillo, Antimo and Salzano, Antonio
                       and Prota, Andrea and Jacobsson, Richard",
      title         = "{The structural design of the decay volume for the Search
                       for Hidden Particles (SHIP) project}",
      journal       = "Archives of Civil and Mechanical Engineering",
      volume        = "21",
      number        = "1",
      pages         = "3",
      year          = "2020",
      url           = "https://cds.cern.ch/record/2812714",
      doi           = "10.1007/s43452-020-00152-9",
}

@article{Gorbunov:2023lga,
    author = "Gorbunov, Dmitry and Kriukova, Ekaterina and Teryaev, Oleg",
    title = "{Scalar decay into pions via Higgs portal}",
    eprint = "2303.12847",
    archivePrefix = "arXiv",
    primaryClass = "hep-ph",
    reportNumber = "INR-TH-2023-003",
    month = "3",
    year = "2023"
}

@article{Planck:2018jri,
    author = "Akrami, Y. and others",
    collaboration = "Planck",
    title = "{Planck 2018 results. X. Constraints on inflation}",
    eprint = "1807.06211",
    archivePrefix = "arXiv",
    primaryClass = "astro-ph.CO",
    doi = "10.1051/0004-6361/201833887",
    journal = "Astron. Astrophys.",
    volume = "641",
    pages = "A10",
    year = "2020"
}

@article{Dai:2014jja,
    author = "Dai, Liang and Kamionkowski, Marc and Wang, Junpu",
    title = "{Reheating constraints to inflationary models}",
    eprint = "1404.6704",
    archivePrefix = "arXiv",
    primaryClass = "astro-ph.CO",
    doi = "10.1103/PhysRevLett.113.041302",
    journal = "Phys. Rev. Lett.",
    volume = "113",
    pages = "041302",
    year = "2014"
}

@article{Cook:2015vqa,
    author = "Cook, Jessica L. and Dimastrogiovanni, Emanuela and Easson, Damien A. and Krauss, Lawrence M.",
    title = "{Reheating predictions in single field inflation}",
    eprint = "1502.04673",
    archivePrefix = "arXiv",
    primaryClass = "astro-ph.CO",
    doi = "10.1088/1475-7516/2015/04/047",
    journal = "JCAP",
    volume = "04",
    pages = "047",
    year = "2015"
}

@article{Caprini:2018mtu,
    author = "Caprini, Chiara and Figueroa, Daniel G.",
    title = "{Cosmological Backgrounds of Gravitational Waves}",
    eprint = "1801.04268",
    archivePrefix = "arXiv",
    primaryClass = "astro-ph.CO",
    doi = "10.1088/1361-6382/aac608",
    journal = "Class. Quant. Grav.",
    volume = "35",
    number = "16",
    pages = "163001",
    year = "2018"
}

@article{ParticleDataGroup:2022pth,
    author = "Workman, R. L. and others",
    collaboration = "Particle Data Group",
    title = "{Review of Particle Physics}",
    doi = "10.1093/ptep/ptac097",
    journal = "PTEP",
    volume = "2022",
    pages = "083C01",
    year = "2022"
}

@article{Planck:2018vyg,
    author = "Aghanim, N. and others",
    collaboration = "Planck",
    title = "{Planck 2018 results. VI. Cosmological parameters}",
    eprint = "1807.06209",
    archivePrefix = "arXiv",
    primaryClass = "astro-ph.CO",
    doi = "10.1051/0004-6361/201833910",
    journal = "Astron. Astrophys.",
    volume = "641",
    pages = "A6",
    year = "2020",
    note = "[Erratum: Astron.Astrophys. 652, C4 (2021)]"
}

@article{Marzola:2016xgb,
    author = "Marzola, Luca and Racioppi, Antonio",
    title = "{Minimal but non-minimal inflation and electroweak symmetry breaking}",
    eprint = "1606.06887",
    archivePrefix = "arXiv",
    primaryClass = "hep-ph",
    doi = "10.1088/1475-7516/2016/10/010",
    journal = "JCAP",
    volume = "10",
    pages = "010",
    year = "2016"
}

@article{Shaposhnikov:2006xi,
    author = "Shaposhnikov, Mikhail and Tkachev, Igor",
    title = "{The nuMSM, inflation, and dark matter}",
    eprint = "hep-ph/0604236",
    archivePrefix = "arXiv",
    reportNumber = "CERN-PH-TH-2006-069",
    doi = "10.1016/j.physletb.2006.06.063",
    journal = "Phys. Lett. B",
    volume = "639",
    pages = "414--417",
    year = "2006"
}

@article{Bezrukov:2009yw,
    author = "Bezrukov, F. and Gorbunov, D.",
    title = "{Light inflaton Hunter's Guide}",
    eprint = "0912.0390",
    archivePrefix = "arXiv",
    primaryClass = "hep-ph",
    doi = "10.1007/JHEP05(2010)010",
    journal = "JHEP",
    volume = "05",
    pages = "010",
    year = "2010"
}

@article{Bezrukov:2020jmo,
    author = "Bezrukov, Fedor and Keats, Abigail",
    title = "{Heavy Light Inflaton and Dark Matter Production}",
    eprint = "2010.06358",
    archivePrefix = "arXiv",
    primaryClass = "hep-ph",
    reportNumber = "MAN/HEP/2020/010",
    doi = "10.1103/PhysRevD.102.115011",
    journal = "Phys. Rev. D",
    volume = "102",
    pages = "115011",
    year = "2020"
}

@article{Guth:1980zm,
    author = "Guth, Alan H.",
    editor = "Fang, Li-Zhi and Ruffini, R.",
    title = "{The Inflationary Universe: A Possible Solution to the Horizon and Flatness Problems}",
    reportNumber = "SLAC-PUB-2576",
    doi = "10.1103/PhysRevD.23.347",
    journal = "Phys. Rev. D",
    volume = "23",
    pages = "347--356",
    year = "1981"
}

@article{Linde:1981mu,
    author = "Linde, Andrei D.",
    editor = "Fang, Li-Zhi and Ruffini, R.",
    title = "{A New Inflationary Universe Scenario: A Possible Solution of the Horizon, Flatness, Homogeneity, Isotropy and Primordial Monopole Problems}",
    reportNumber = "LEBEDEV-81-229",
    doi = "10.1016/0370-2693(82)91219-9",
    journal = "Phys. Lett. B",
    volume = "108",
    pages = "389--393",
    year = "1982"
}

@article{Bramante:2016yju,
    author = "Bramante, Joseph and Cook, Jessica and Delgado, Antonio and Martin, Adam",
    title = "{Low Scale Inflation at High Energy Colliders and Meson Factories}",
    eprint = "1608.08625",
    archivePrefix = "arXiv",
    primaryClass = "hep-ph",
    doi = "10.1103/PhysRevD.94.115012",
    journal = "Phys. Rev. D",
    volume = "94",
    number = "11",
    pages = "115012",
    year = "2016"
}

@article{Barman:2024nhr,
    author = "Barman, Basabendu and Bhattacharya, Subhaditya and Jahedi, Sahabub and Pradhan, Dipankar and Sarkar, Abhik",
    title = "{Lepton collider as a window to reheating via freezing in dark matter detection. Part I}",
    eprint = "2406.11963",
    archivePrefix = "arXiv",
    primaryClass = "hep-ph",
    doi = "10.1016/j.physletb.2025.139863",
    journal = "Phys. Lett. B",
    volume = "869",
    pages = "139863",
    year = "2025"
}

@article{Barman:2024tjt,
    author = "Barman, Basabendu and Bhattacharya, Subhaditya and Jahedi, Sahabub and Pradhan, Dipankar and Sarkar, Abhik",
    title = "{Lepton collider as a window to reheating via freezing in dark matter detection. Part II}",
    eprint = "2410.18198",
    archivePrefix = "arXiv",
    primaryClass = "hep-ph",
    doi = "10.1007/JHEP07(2025)157",
    journal = "JHEP",
    volume = "07",
    pages = "157",
    year = "2025"
}

@article{Pradhan:2026maz,
    author = "Pradhan, Dipankar and Mondal, Niloy and Sarkar, Abhik and Ghosh, Anupam and Sharma, Shashwat and Thomas Arun, Mathew and Barman, Basabendu",
    title = "{From WIMP to FIMP during reheating: collider vs non-collider probes for p-wave annihilation}",
    eprint = "2605.27521",
    archivePrefix = "arXiv",
    primaryClass = "hep-ph",
    month = "5",
    year = "2026"
}

@article{Okada:2019opp,
    author = "Okada, Nobuchika and Raut, Digesh",
    title = "{Hunting inflatons at FASER}",
    eprint = "1910.09663",
    archivePrefix = "arXiv",
    primaryClass = "hep-ph",
    doi = "10.1103/PhysRevD.103.055022",
    journal = "Phys. Rev. D",
    volume = "103",
    number = "5",
    pages = "055022",
    year = "2021"
}

@article{Ema:2017ckf,
    author = "Ema, Yohei and Karciauskas, Mindaugas and Lebedev, Oleg and Rusak, Stanislav and Zatta, Marco",
    title = "{Higgs{\textendash}inflaton mixing and vacuum stability}",
    eprint = "1711.10554",
    archivePrefix = "arXiv",
    primaryClass = "hep-ph",
    doi = "10.1016/j.physletb.2018.10.074",
    journal = "Phys. Lett. B",
    volume = "789",
    pages = "373--377",
    year = "2019"
}

@article{AtacamaCosmologyTelescope:2025blo,
    author = "Louis, Thibaut and others",
    collaboration = "Atacama Cosmology Telescope",
    title = "{The Atacama Cosmology Telescope: DR6 power spectra, likelihoods and {\ensuremath{\Lambda}}CDM parameters}",
    eprint = "2503.14452",
    archivePrefix = "arXiv",
    primaryClass = "astro-ph.CO",
    reportNumber = "FERMILAB-PUB-25-0071-PPD",
    doi = "10.1088/1475-7516/2025/11/062",
    journal = "JCAP",
    volume = "11",
    pages = "062",
    year = "2025"
}

@article{AtacamaCosmologyTelescope:2025nti,
    author = "Calabrese, Erminia and others",
    collaboration = "Atacama Cosmology Telescope",
    title = "{The Atacama Cosmology Telescope: DR6 constraints on extended cosmological models}",
    eprint = "2503.14454",
    archivePrefix = "arXiv",
    primaryClass = "astro-ph.CO",
    reportNumber = "FERMILAB-PUB-25-0157-PPD",
    doi = "10.1088/1475-7516/2025/11/063",
    journal = "JCAP",
    volume = "11",
    pages = "063",
    year = "2025"
}

@article{Bezrukov:2014nza,
    author = "Bezrukov, F. and Gorbunov, D.",
    title = "{Relic Gravity Waves and 7 keV Dark Matter from a GeV scale inflaton}",
    eprint = "1403.4638",
    archivePrefix = "arXiv",
    primaryClass = "hep-ph",
    doi = "10.1016/j.physletb.2014.07.060",
    journal = "Phys. Lett. B",
    volume = "736",
    pages = "494--498",
    year = "2014"
}

@article{Racioppi:2018zoy,
    author = "Racioppi, Antonio",
    title = "{New universal attractor in nonminimally coupled gravity: Linear inflation}",
    eprint = "1801.08810",
    archivePrefix = "arXiv",
    primaryClass = "astro-ph.CO",
    doi = "10.1103/PhysRevD.97.123514",
    journal = "Phys. Rev. D",
    volume = "97",
    number = "12",
    pages = "123514",
    year = "2018"
}

@article{Gialamas:2025kef,
    author = "Gialamas, Ioannis D. and Karam, Alexandros and Racioppi, Antonio and Raidal, Martti",
    title = "{Has ACT measured radiative corrections to the tree-level Higgs-like inflation?}",
    eprint = "2504.06002",
    archivePrefix = "arXiv",
    primaryClass = "astro-ph.CO",
    doi = "10.1103/6fpc-67s1",
    journal = "Phys. Rev. D",
    volume = "112",
    number = "10",
    pages = "103544",
    year = "2025"
}

@article{Wolf:2025ecy,
    author = "Wolf, William J.",
    title = "{Inflationary attractors and radiative corrections in light of ACT data}",
    eprint = "2506.12436",
    archivePrefix = "arXiv",
    primaryClass = "astro-ph.CO",
    doi = "10.1088/1475-7516/2026/02/088",
    journal = "JCAP",
    volume = "02",
    pages = "088",
    year = "2026"
}

@article{Lebedev:2023zgw,
    author = "Lebedev, Oleg and Mambrini, Yann and Yoon, Jong-Hyun",
    title = "{On unitarity in singlet inflation with a non-minimal coupling to gravity}",
    eprint = "2305.05682",
    archivePrefix = "arXiv",
    primaryClass = "hep-ph",
    doi = "10.1088/1475-7516/2023/08/009",
    journal = "JCAP",
    volume = "08",
    pages = "009",
    year = "2023"
}

@article{Burgess:2009ea,
    author = "Burgess, C. P. and Lee, Hyun Min and Trott, Michael",
    title = "{Power-counting and the Validity of the Classical Approximation During Inflation}",
    eprint = "0902.4465",
    archivePrefix = "arXiv",
    primaryClass = "hep-ph",
    reportNumber = "PI-PARTPHYS-121",
    doi = "10.1088/1126-6708/2009/09/103",
    journal = "JHEP",
    volume = "09",
    pages = "103",
    year = "2009"
}

@article{Barbon:2009ya,
    author = "Barbon, J. L. F. and Espinosa, J. R.",
    title = "{On the Naturalness of Higgs Inflation}",
    eprint = "0903.0355",
    archivePrefix = "arXiv",
    primaryClass = "hep-ph",
    reportNumber = "IFT-UAM-CSIC-09-10, UAB-FT-665",
    doi = "10.1103/PhysRevD.79.081302",
    journal = "Phys. Rev. D",
    volume = "79",
    pages = "081302",
    year = "2009"
}

@article{Feng:2022inv,
    author = "Feng, Jonathan L. and others",
    title = "{The Forward Physics Facility at the High-Luminosity LHC}",
    eprint = "2203.05090",
    archivePrefix = "arXiv",
    primaryClass = "hep-ex",
    reportNumber = "UCI-TR-2022-01, CERN-PBC-Notes-2022-001, INT-PUB-22-006, BONN-TH-2022-04, FERMILAB-PUB-22-094-ND-SCD-T",
    doi = "10.1088/1361-6471/ac865e",
    journal = "J. Phys. G",
    volume = "50",
    number = "3",
    pages = "030501",
    year = "2023"
}

@inproceedings{FPFWorkingGroups:2025rsc,
    author = "Anchordoqui, Luis A. and others",
    collaboration = "FPF Working Groups",
    title = "{The Forward Physics Facility at the Large Hadron Collider}",
    eprint = "2503.19010",
    archivePrefix = "arXiv",
    primaryClass = "hep-ex",
    month = "3",
    year = "2025"
}

@article{Bostan:2019fvk,
    author = "Bostan, Nilay and {\c{S}}eno{\u{g}}uz, Vedat Nefer",
    title = "{Quartic inflation and radiative corrections with non-minimal coupling}",
    eprint = "1907.06215",
    archivePrefix = "arXiv",
    primaryClass = "astro-ph.CO",
    doi = "10.1088/1475-7516/2019/10/028",
    journal = "JCAP",
    volume = "10",
    pages = "028",
    year = "2019"
}

@article{Allison:2014hna,
    author = "Allison, Kyle and Hill, Christopher T. and Ross, Graham G.",
    title = "{An ultra-weak sector, the strong CP problem and the pseudo-Goldstone dilaton}",
    eprint = "1409.4029",
    archivePrefix = "arXiv",
    primaryClass = "hep-ph",
    reportNumber = "FERMILAB-PUB-14-414-T",
    doi = "10.1016/j.nuclphysb.2014.12.022",
    journal = "Nucl. Phys. B",
    volume = "891",
    pages = "613--626",
    year = "2015"
}

@article{Allison:2014zya,
    author = "Allison, Kyle and Hill, Christopher T. and Ross, Graham G.",
    title = "{Ultra-Weak Sector, Higgs Boson Mass, and the Dilaton}",
    eprint = "1404.6268",
    archivePrefix = "arXiv",
    primaryClass = "hep-ph",
    reportNumber = "FERMILAB-PUB-14-086-T",
    doi = "10.1016/j.physletb.2014.09.041",
    journal = "Phys. Lett. B",
    volume = "738",
    pages = "191--195",
    year = "2014"
}

@article{Ballesteros:2015noa,
    author = "Ballesteros, Guillermo and Tamarit, Carlos",
    title = "{Radiative plateau inflation}",
    eprint = "1510.05669",
    archivePrefix = "arXiv",
    primaryClass = "hep-ph",
    reportNumber = "SACLAY-T15-176, CERN-PH-TH-2015-247, IFT-UAM-CSIC-15-11, DCPT-15-116, IPPP-15-58",
    doi = "10.1007/JHEP02(2016)153",
    journal = "JHEP",
    volume = "02",
    pages = "153",
    year = "2016"
}

@article{Oda:2017zul,
    author = "Oda, Satsuki and Okada, Nobuchika and Raut, Digesh and Takahashi, Dai-suke",
    title = "{Nonminimal quartic inflation in classically conformal U(1)$_X$ extended standard model}",
    eprint = "1711.09850",
    archivePrefix = "arXiv",
    primaryClass = "hep-ph",
    doi = "10.1103/PhysRevD.97.055001",
    journal = "Phys. Rev. D",
    volume = "97",
    number = "5",
    pages = "055001",
    year = "2018"
}
\bibliographystyle{JHEP}
\end{document}